\documentclass[final,leqno]{siamltex}
\usepackage[dvips]{graphicx}
\usepackage{subfigure}
\usepackage{amsmath}
\usepackage{amsfonts}

\usepackage{mathtools}
\usepackage{booktabs}
\usepackage{color}

\newtheorem{thm}{Theorem}

\newtheorem{prop}{Proposistion}
\newtheorem{lem}{Lemma}
\newtheorem{cor}{Corollary}
\newtheorem{asmp}{Assumption}

\newtheorem{rem}{Remark}

\newtheorem{exam}{Example}

\def \R{\mathbb{R}}
\def \F{\mathcal{F}}

\def \A{\mathcal{A}}

\def \P{\mathcal{P}}
\def \mP{\mathbb{P}}
\def \VaR{ \textrm{VaR}}

\def \E {\textrm{E}}
\def \CVaR {\textrm{CVaR}}

\def \1{\mathbf{1}}
\def \L{\mathcal{L}}
\def \B{\mathcal{B}}

\def \O{\mathcal{O}}
\def \M{\mathcal{M}}

\def \pp{\partial}

\def \X{\mathcal{X}}

\title{DYNAMIC MEAN-LPM and MEAN-CVaR PORTFOLIO OPTIMIZATION IN CONTINUOUS-TIME \thanks{ This research work was partially supported by
Natural Science Foundation of China under grant 71201102, by Ph.D. Programs Foundation of Ministry of Education of China under grand 20120073120037,
and by Hong Kong Research Grants Council under grants CUHK 414513 and
CUHK414610. The third author is grateful to the support from Patrick Huen Wing Ming Professorship of Systems Engineering \& Engineering Management.}}

\author{Jianjun Gao\thanks{Department of Automation, Shanghai Jiao Tong University, Shanghai, China.}
\and Ke Zhou \thanks{Department of Systems Engineering and Engineering Management, The Chinese University of Hong Kong, Hong Kong.}
\and Duan LI\thanks{Corresponding author. Department of Systems Engineering and Engineering Management, The Chinese University of Hong Kong, Hong Kong. E-mail: dli@se.cuhk.edu.hk.}
\and Xiren Cao \thanks{Department of Automation, Shanghai Jiao Tong University, Shanghai, China.}
}

\begin{document}
\maketitle
\newcommand{\slugmaster}{%
\slugger{siads}{xxxx}{xx}{x}{x--x}}%slugger should be set to juq, siads, sifin, or siims

\begin{abstract}
Instead of controlling ``symmetric'' risks measured by central moments of investment return or terminal wealth, more and more portfolio models have shifted their focus to manage ``asymmetric'' downside risks that the investment return is below certain threshold. Among the existing downside risk measures, the lower-partial moments (LPM) and conditional value-at-risk (CVaR) are probably most promising. In this paper we investigate the dynamic mean-LPM and mean-CVaR portfolio optimization problems in continuous-time, while the current literature has only witnessed their static versions. Our contributions are two-fold, in both building up tractable formulations and deriving corresponding analytical solutions. By imposing a limit funding level on the terminal wealth, we conquer the ill-posedness exhibited in the class of mean-downside risk portfolio models. The limit funding level not only enables us to solve both dynamic mean-LPM and mean-CVaR portfolio optimization problems, but also offers a flexibility to tame the aggressiveness of the portfolio policies generated from such mean - downside risk models. More specifically,  for a general market setting, we prove the existence and uniqueness of the Lagrangian multiplies, which is a key step in applying the martingale approach, and establish a theoretical foundation for developing efficient numerical solution approaches. Moreover, for situations where the opportunity set of the market setting is deterministic, we derive analytical portfolio policies for both dynamic mean-LPM and mean-CVaR formulations.
\end{abstract}



\pagestyle{myheadings}
\thispagestyle{plain}
\markboth{J.J. Gao, K. Zhou, D. Li, X.R. Cao}{DYNAMIC MEAN-LPM and MEAN-CVaR PORTFOLIO SELECTION}

%****************************************************************************
\section{Introduction}
The mean-variance (MV) formulation pioneered by Markowitz \cite{Markowitz:1952} sixty years ago has laid the foundation of modern portfolio theory. Most importantly, the mean-variance model captures the essential multiobjective nature between the two conflicting goals in portfolio selection, i.e., between maximizing the investment return and minimizing the investment risk. As a natural generalization of the mean-variance analysis, the framework of mean-risk trade-off analysis has become a standard in portfolio management. Under the framework of mean-risk trade-off analysis, a risk measure always serves a purpose to map investment uncertainty to a quantitative level such that trade-off can be computed explicitly against the expected investment return. Such a straightforward appealing approach of risk management is in general more favored by both practitioners in financial industry and researchers in academic field, when compared to the more abstract, albeit more mathematically rigorous, expected utility maximization framework. However, selecting an appropriate risk measure is essentially not only a science, but also an art.

While the variance term penalizes uncertainties on both sides of the mean, numerous downside risk measures have been proposed in the last half century to quantify the risk that the investment return is below certain target. Among these downside risk measures, the lower-partial moments (LPM) proposed by Fishbburn \cite{Fishburn:1977} form one most important class with prominent features. The LPM enables us to represent a general form of downside risk measures with two parameters, the benchmark level $\gamma$, which is set by the investor himself, and the order of the moments, $q$, which represents the risk attitude of the investor. Due to the freedom offered by different combinations of the pair $q$ and $\gamma$, we can adopt LPM to pursue different investment goals in portfolio optimization. For example, setting $q=0$ in LMP yields the shortfall probability, which is also equivalent to the safety-first rule proposed by Roy \cite{Roy:1952}; Setting $q=1$ gives rise to the risk measure of the expected regret (ER) (see Dembo \cite{Dembo:1999}); and setting $q=2$ leads to the risk measure of semideviation below the target, or the semivariance if $\gamma$ is set as the expected terminal wealth. Bawa and Lindenberg \cite{Bawa:1977} show that LPMs associated with $q=0$, 1, or 2  correspond to the first, second or third degree stochastic dominance, respectively. Compared with the variance, the LPM is more consistent with the classical utility theory and the rule of stochastic dominance (see e.g., \cite{Ogryczak:2002}). Konno et al. \cite{Konno:2002} demonstrate the prominence of LPM in the practice of portfolio management via empirical tests. Zhu et al. further consider robust portfolio selection under LPM risk measures \cite{ZhuLiWang:2009}.

The Value-at-Risk (VaR), defined as the threshold point with a specified exceeding probability of great loss, becomes popular in the financial industry since the mid 90s. However, the VaR has been widely criticized for some of its undesired properties. More specifically, VaR fails to satisfy the axiomatic system of coherent risk measures proposed by Artzner et al. \cite{Artzner:1999}.  Most critically, the non-convexity of VaR leads to some difficulty in solving the corresponding portfolio optimization problem. On the other hand, the conditional Value-at-Risk (CVaR), also known as the expected shortfall, is defined as the expected value of the loss exceeding the VaR \cite{Rockafellar:2000}. CVaR possesses several good properties, such as convexity, monotonicity and homogeneity. Rockafellar and Uryasev \cite{Rockafellar:2000} \cite{Rockafellar:2002} prove that CVaR can be computed by solving an auxiliary linear programming problem in which the VaR needs not to be known in advance. After the fundamental work of Rockafellar and Uryasev (\cite{Rockafellar:2000} \cite{Rockafellar:2002}), CVaR has been widely applied in various applications of portfolio selection and risk management, e.g., derivative portfolio \cite{Alexander:2006},  credit risk optimization \cite{Andersson:2001}, and robust portfolio management \cite{Zhu:2009}.

Almost all the mean-downside risk portfolio optimization models studied in the above literature have been confined to static settings, from which the derived portfolio policy is of a buy-and-hold nature. Without a doubt, such a class of static models is not suitable for investment problems with a long investment horizon. The past decade has witnessed some research works that investigate mean-CVaR portfolio optimization using stochastic programming approach \cite{Fabian:2006} \cite{Fabian:2007} \cite{Hibibi:2006}. As stochastic programming formulations adopt both discrete time and discrete state in their model settings, this kind of models with discrete states suffers from a heavy computational burden, and can only deal with two - or three - stage problems. Within dynamic mean-risk portfolio optimization models, the most matured development seems to lie in the subject of dynamic mean-variance (MV) portfolio optimization. Although the mean-variance analysis starts the area of portfolio selection, its extension to a dynamic MV version has  been blocked for almost four decades, due to the nonseparability of the variance term in the sense of dynamic programming. After Li and Ng \cite{LiNg:2000} and Zhou and Li \cite{ZhouLi:2000} derive the explicit portfolio policies, respectively, for discrete-time and continuous-time MV portfolio selection formulations, by using the embedding scheme, the dynamic MV models has been developed by leaps and bounds, see, for examples, \cite{LiZhou:2001} \cite{Lim:2002} \cite{Zhu:2004}  \cite{Jin:2008}. Recently, the subject of time consistency in dynamic MV portfolio optimization has been attracting increasing attention  (see, e.g., \cite{Basak:2010} \cite{CLWZ:2011} \cite{Bjork:2012}). Although the mean-downside risk models seem to be a natural extension of dynamic MV models, Jin et al. \cite{Jin:2005} show that a general class of mean-downside risk portfolio optimization models under a continuous-time setting is ill-posed in the sense that the optimal value cannot be achieved. Besides such a negative result, there do exist some research works related to the continuous-time portfolio selection problems in which  the  downside-risk measure plays a role. For example, Basak and Shapiro \cite{Basak:2001} consider the continuous-time utility maximization model with a VaR constraint. By using the stochastic control approach, Yiu \cite{Yiu:2004} study a problem similar to \cite{Basak:2001}. However, the VaR risk constraint in Yiu \cite{Yiu:2004} is defined over the entire investment process. Gundel and Weber \cite{Gundel:2008} extend the VaR risk constraint to a shortfall risk constraint. Recently, Chiu et al. \cite{Chiu:2012} solve the dynamic asset-liability management problem under the safety-first criteria, which can be regarded as the shortfall probability measure.

We consider in this paper the mean-downside risk portfolio optimization problem in a continuous-time setting. More specifically, we investigate both the dynamic mean-LPM and mean-CVaR portfolio optimization problems. In recognizing the ill-posedness of such problems (see, e.g., Jin et al. \cite{Jin:2005}), we adopt a similar solution idea as in \cite{Chiu:2012} to attach to this class of problems an upper limit on the funding level of the terminal wealth. In the continuous-time mean-LPM and mean-CVaR portfolio optimization models, if the terminal wealth is unlimited, the investor will act extremely aggressively to push his terminal wealth to the infinity. Adding a limit on the funding level will tame such an irregular portfolio policy to a reasonable level. Thus, such an upper bound can be also regarded as a designing variable to control the aggressiveness level of the investor. We further prove that the probability that the terminal wealth reaches such an upper bound is decreasing with respect to the magnitude of the upper level. For general market opportunity set, we prove the exitance and uniqueness of Lagrangian multipliers, which is the key step to apply the martingale approach. These theoretical results pave a foundation to develop numerical solution schemes to solve dynamic mean-LPM and mean-CVaR portfolio optimization problems. When the market opportunity set is deterministic, we further derive semi-analytical portfolio policies for both the mean-LMP and mean-CVaR portfolio optimization problems. The dynamic mean-LPM portfolio policy demonstrates very distinct features when compared with the dynamic MV portfolio policy. When the market condition is good, the mean-LPM investor tends to invest more aggressively in the risky assets when compared to an MV investor. When the market condition is in the medium state, the mean-LPM investor prefers to allocate more wealth in the risk-free asset. However, when the market condition is in a bad state, the mean-LPM investor allocates again  more wealth in the risky assets than the MV investor. This phenomena can be regarded as the gambling effect of dynamic mean-LPM investors. In summary, the mean-LPM investment policies show a feature of a  two-side threshold type, i.e., at any time $t$, when the current wealth is, respectively, below or above certain levels, the investor increases his allocations in the risky assets. As for the dynamic mean-CVaR portfolio policy, our experiment result with real market data shows that the CVaR measure can be improved significantly when compared with the buy-and-hold mean-CVaR portfolio policy of a static type.

The remaining of the paper is organized as follows. We present the market setting and the dynamic mean-LMP and dynamic mean-CVaR portfolio optimization problem formulations in Section \ref{se_formulation}. We derive the optimal portfolio policies for the dynamic mean-LPM and dynamic mean-CVaR optimization problems in Section \ref{se_LPM} and \ref{se_cvar}, respectively. We then present illustrative examples to compare the dynamic mean-LPM portfolio policy with the dynamic mean-variance portfolio policy and the dynamic mean-CVaR portfolio policy with the static portfolio policy in Section \ref{se_example}. Finally, we conclude our paper in Section \ref{Conclusion}. Throughout the entire paper, notation $\mathbf{1}_\mathcal{B}$  denotes the indicator function, i.e., $\1_{\mathcal{B}}=1$ if the condition $\mathcal{B}$ holds true and $\mathbf{1}_{\mathcal{B}}=0$, otherwise; $A^{\prime}$ denotes the transpose of matrix $A$, and $(a)_+$ denotes the nonnegative part of $a$, i.e., $(a)_+= a \mathbf{1}_{a\geq 0}$. To simplify our notations, we use $(a)_+^q$ for $((a)_+)^q$ which means the $q$-th power function of $(a)_+$. Finally, the cumulative distribution of the standard normal random variable $X$ is denoted by $\Phi(y):=\mathbb{P}(X\leq y)=\frac{1}{\sqrt{2\pi}}\int_{-\infty}^y \exp(-\frac{s^2}{2})ds$.

%*****************************************************************************
\section{Market Setting and Problem Formulations}\label{se_formulation}
We consider a market with $n$ risky assets and one risk free asset which can be traded continuously within time horizon $[0,T]$. All the randomness are modeled by a complete filtrated probability space $(\Omega, \F, \mP, \{\F_t\}_{t\geq 0})$, on which  an $\F_t$ adapted $n$-dimensional Brownian motion $W(t)=\left(W_1(t),\cdots, W_n(t)\right)^{\prime}$ is defined, where $W_i(t)$ and $W_j(t)$ are mutually independent for all $i\not=j$. Let $\L^2_{\F}(0,T;\R^n)$ be the set of $\R^n$-valued, $\F_t$-adapted and square integrable stochastic processes, and $\L^2_{\F_T}(\Omega;\R^n)$ the set of $\R^n$ valued $\F_T$-measurable random variables.

The price process $S_0(t)$ of the risk-free asset is governed by the following ordinary differential equation,
\begin{align}
  \begin{dcases}
    dS_0(t)=r(t)S_0(t)dt, ~~t\in [0,T],\\
    S_0(0)=s_0>0,
  \end{dcases} \label{def_riskfree}
\end{align}
where $r(t)$ is the risk free return rate, which is $\F_t$ measurable scalar-valued stochastic process. The price process of the $n$ risky assets satisfies the following system of stochastic differential equations (SDE):
\begin{align}
  \begin{dcases}
  dS_i(t) = S_i(t)\big(\mu_i(t) dt + \sum_{j=1}^n \sigma_{ij}(t) d W_j(t) \big),~~t\in[0,T],~~i = 1, \ldots, n,\\
  S_i(0) =s_i>0,~~i = 1, \ldots, n,
  \end{dcases}\label{def_risky_sde}
\end{align}
where $\mu_i(\cdot)$ and $\sigma_{ij}(\cdot)$ are the appreciation rate and volatility, respectively. We assume that all $\mu_i(\cdot)$ and $\sigma_{ij}(\cdot)$ are uniformly bounded, scalar-valued $\F_t$-measurable stochastic processes. Furthermore, we assume that the volatility matrix $\sigma(t):=\{\sigma_{ij}(t)\}\mid_{i,j=1}^{n,n}$ satisfies the following nondegeneracy condition,
\begin{align}
\sigma(t)\sigma^{\prime}(t) \succ \epsilon I, ~~\textrm{for ~all}~0\leq t \leq T,~~ a.s.,
\end{align}
for some $\epsilon>0$\footnote{`a.s.' stands for `almost surely', which excludes events with zero occurrence probability. In the following discussion, we simply ignore such a term for the random variables that satisfy certain condition.}. Under the above setting, we have a complete market model for the securities.

An investor with initial wealth $x_0$ enters the market at time $0$ and continuously allocates his wealth  in the $n$ risky assets and the risk-free asset within time horizon $[0,T]$. Let $x(t)$ be the total wealth of the investor at time $t$. Denote the portfolio process by $\pi(t)=\big(\pi_1(t),\cdots, \pi_n(t)\big)^{\prime}$ with $\pi(\cdot)\in \L^2_{\F}(0,T; \R^n)$, where $\pi_i(t)$ is the dollar amount allocated to risky asset $i$ at time $t$. As we do not consider in this research the transaction cost during the investment process, the wealth process of the investor, $x(t)$, then satisfies the following stochastic differential equation (SDE),
\begin{align}
\begin{dcases}
dx(t) = \Big( r(t)x(t)+b(t)^{\prime}\pi(t) \Big)dt + \pi(t)^{\prime}\sigma(t) dW(t), \\
x(0)=x_0,
\end{dcases}
\label{def_wealth}
\end{align}
where $b(t)$ is the excess return defined by
\begin{align*}
b(t):=\left(
                 \begin{array}{cccc}
                   \mu_1(t)-r(t) &  \mu_2(t)-r(t) & \cdots & \mu_n(t)-r(t)
                 \end{array}
               \right)^{\prime}.
\end{align*}
In this research, we focus our investigation on mean-downside risk portfolio optimization. In particularly, we are interested in studying the following mean-LPM model,
\begin{align*}
(\P_{lpm}^{q})~~&~~\min_{\pi(\cdot)\in \L^2_{\F}(0,T; \R^n) }~~\E[(\gamma-x(T) )_+^q ]\\
\textrm{Subject to}~&~\begin{dcases}
                        \E[x(T)]\geq d, \\
                        \textrm{\{$x(\cdot)$,$\pi(\cdot)$\} statisfies (\ref{def_wealth}) }, \\
                        0 \leq x(T) \leq B,
                        %\pi(\cdot) \in \L_{\F}^2(0,T;\R^n),
                      \end{dcases}%\label{def_P1_constraint}
\end{align*}
where $d$ is the minimum expected wealth which the investor would like to attain, $B$ is an upper bound of the attainable final wealth imposed by the investor, $\gamma \in \R$ is a given benchmark level, and $q$ is a given nonnegative integer, which represents the order of the moment. Adopting model $(\P_{lpm}^{q})$ implies that the investor only cares about the scenarios where $x(T)$ is less than the benchmark level $\gamma$, which the investor sets as a threshold for ``disastrous'' terminal wealth. When $q$ = 0, from our notations, we have  $(\gamma-x(T))_+^0=\1_{x(T)\leq \gamma}$ and thus $\E[(\gamma-x(T))_+^0 ]=\mP(x(T)\leq \gamma)$, which is the disaster probability considered by Roy in his pioneering safety-first principle \cite{Roy:1952}, while $\gamma$ can be viewed as the disaster level. When $q$ = 1 and $\gamma$ = $\E[x(T)]$,  the downside risk measure $\E[(\E[x(T)]-x(T))_+^1]$ becomes the semi-absolute deviation
(or the target semi-absolute-deviation). When $q$ = 2 and $\gamma$ = $\E[x(T)]$,  the downside risk measure $\E[( \E[x(T)]-x(T) )_+^2]$ yields the
semi-variance (or the target semi-variance). Let $\bar{x}_T$ be a given safe-level of the terminal wealth. One possible candidate of $\bar{x}_T$ could be
\begin{align}
\bar{x}_T = \E[e^{\int^T_{0}r(s)ds}]x_0, \label{def_barxT}
\end{align}
which is the expected terminal wealth when investing all initial wealth in the risk free account. For the upper bound $B$, we reasonably assume $B>\max\{d, \bar{x}_T, \gamma\}$.

In our work, we also study dynamic mean-CVaR portfolio optimization. We define first the loss of investment as follows,
\begin{align}
f(x(T)):= \bar{x}_T -x(T).\label{def_loss}
\end{align}
We adopt the definition of CVaR by Rockafellar and Uryasev \cite{Rockafellar:2002} for investment loss and use the notation $\CVaR[f(x(T))]$ to denote the CVaR of the investment loss. The mean-CVaR portfolio optimization model is now formally posted as follows,
\begin{align*}
(\P_{cvar})~&~\min_{\pi(\cdot)\in \L^2_{\F}(0,T; \R^n)}~ \CVaR[f(x(T))],\\%\label{def_P1_obj}\\
\textrm{Subject to}~&~
                      \begin{dcases}
                        \E[x(T)]\geq d, \\
                        \textrm{\{$x(\cdot)$,$\pi(\cdot)$\} statisfies (\ref{def_wealth})}, \\
                        0 \leq x(T) \leq B,\\
                        %\pi(\cdot) \in \L_{\F}^2(0,T;\R^n),
                      \end{dcases}%\label{def_P1_constraint}
\end{align*}
where all the other notations are defined the same as in $(\P_{lpm}^{q})$.

As we will demonstrate later in this paper, the upper bound, $B$, imposed on the terminal wealth essentially controls the aggressiveness of the portfolio policy. The larger the value of $B$, the more aggressive the portfolio policy becomes. If we let $B$ go to infinite, both problems $(\P_{lpm}^{q})$ and $(\P_{cvar})$ will become ill-posed (see, e.g., \cite{Jin:2005}), i.e., the investor would take an infinite position. From the view point of real applications, any portfolio that generates extremely high level of terminal wealth is not realistic. Thus, imposing an upper bound on the terminal wealth, as proposed in \cite{Chiu:2012}, is reasonable and justifiable. Furthermore, such an upper bound can be also regarded as a designing variable to control the aggressiveness level of the investor. We also prove that the probability that the terminal wealth reaches its upper bound is monotonically decreasing with respect to the level of the upper bound. Thus, a formulation with a very large upper bound can be regarded as an approximation to the formulation without an upper bound. Note also that the no-bankruptcy constraint at the terminal time, $x(T)\geq 0$, actually ensures no-bankruptcy for the entire wealth process, i.e., $x(t)\geq 0$, for $t\in [0,T]$ (see Proposition 2.1 in \cite{Bielecki:2005}).

%******************************************************************************
\section{Optimal Portfolio Policy For Dynamic Mean-LPM Formulation}\label{se_LPM}

We develop in this section a solution scheme for problem $(\P_{lpm}^{q})$ using the martingale approach (see, for examples, \cite{Pliska:1986} and \cite{Karatzas:1998}). The main idea of the martingale approach is to find first the optimal terminal wealth $x^*(T)$ by solving a static optimization problem and to identify then the optimal portfolio policy $\pi^*(\cdot)$ process to replicate (generate) such an optimal wealth distribution of $x^*(T)$.

\subsection{Optimal terminal wealth}
From our {\it complete} market setting in (\ref{def_risky_sde}), we can find a unique equivalent martingale measure (EMM) such that the discounted price processes of the risk assets are martingale. Let the Radon-Nikod\'{y}m derivative of the EMM, $\tilde{\mP}$, with respect to the original measure $\mP$ be $\xi$, i.e., $\xi := d \tilde{\mP}/d \mP$, where $\xi$ is an $\F_T$-measurable random variable. From the Girsanov Theorem \cite{Karatzas:1998}, the Radon-Nikod\'{y}m derivative process $\xi(t)=\E[\xi |\F_t]$ can be expressed as the exponential martingale, $d\xi(t)=\xi(t)\theta(t)^{\prime}dW(t)$, where $\theta(t)$ is $m\times 1$ vector-valued $\F_t$-adapted stochastic process vector such that the choice of $\theta(t)$ makes the process $d\tilde{W}(t)=\theta(t)dt+dW(t)$ to be the Brownian motion under probability $\tilde{\mP}$. To eliminate the drift term of the discount price process of the securities, we let $\theta(t)$ be
\begin{align}
\theta(t)=\sigma(t)^{-1}b(t), a.s., \textrm{for} ~t\in [0,T]. \label{def_theta}
\end{align}
Then, we define the state price density as
$z(t):= \xi(t)/S_0(t)$ which satisfies the following SDE,
\begin{align}
  \begin{dcases}
    d z(t)=-z(t)\big( r(t)dt + \theta(t)^{\prime}dW(t) \big), \\
    z(0)=1,
  \end{dcases}\label{def_z}
\end{align}
or, equivalently, we can express $z(t)$ as
\begin{align*}
z(t)=\exp\left\{-\int^{t}_0\big(r(s)+\frac{1}{2}\|\theta(s)\|^2\big) ds-\int^t_0 \theta(s)^{\prime}dW(s)  \right\}.
\end{align*}
In the literature, $z(t)$ is also referred as the \textit{deflator process}, which transfers the discounted wealth process $x(t)$ to a martingale, i.e., we have
\begin{align*}
z(t)x(t)=\E[z(s)x(s)~|~\F_t],
\end{align*}
for any $t<s\leq T$.  By using such a property, the optimal terminal wealth $x(T)$ of the problem $({\P}_{lpm}^{q})$ can be found by solving the following static optimization problem,
\begin{align}
(\A^q)~&~~\min_{x(T)\in \L_{\F_T}^2(\Omega,\R) }~\E\big[(\gamma-x(T))_+^q  \big], \notag\\
\textrm{Subject to}~&~\E[x(T)]\geq d, \label{A_lpm_cnst1}\\
~&~\E[z(T)x(T)]=x_0, \label{A_lpm_cnst2} \\
~&~0 \leq x(T) \leq B. \notag
\end{align}
Before we solve problem ($\A^q$) we need the following lemma.

\begin{lem}\label{lem_lag}
Given the following two problems ($\mathcal{B}$) and ($\mathcal{L}$),
\begin{align*}
(\mathcal{B})~&~~\min_{Y \in C}~\E[f(Y)],\notag\\
\textrm{Subject to:}~&~~ \E[Y]-b\geq 0,\\
~&~~\E[ZY]-a=0,
\end{align*}
and
\begin{align*}
(\mathcal{L}(\lambda_1,\lambda_2))&~~\min_{Y\in C}~\E[f(Y)-\lambda_1(Y-b)+\lambda_2(ZY-a)],
\end{align*}
where $C\subset \L^2_{\F_T}(\Omega;\R)$ is a convex set, $f(\cdot)$ is a scalar-valued convex function, $Z\in\L^2_{\F_T}(\Omega;\R)$ is a random variable, $a\in \R$, $b\in \R$, $\lambda_1 \in \R_+$ and $\lambda_2 \in \R$. If $Y^*$ solves problem $(\mathcal{L}(\lambda_1^*,\lambda_2^*))$ for some $\lambda_1^*$ and $\lambda_2^*$ and satisfies $\E[Y^*]\geq b$ and $\E[ZY^*]=a$, then $Y^*$ solves problem $(\mathcal{B})$ with $\lambda_1^*(\E[Y^*]-b)=0$ and $\lambda_2^*(\E[Z Y^*]-a)=0$. On the other hand, if problem $(\B)$ has a solution $Y^*$, then there exist $\lambda_1^*$ and $\lambda_2^*$ such that $Y^*$ also solves problem $(\L(\lambda^*_1,\lambda^*_2))$.
\end{lem}

We place the proof of Lemma \ref{lem_lag} in the Appendix. Lemma \ref{lem_lag} basically shows that problem $(\B)$ can be solved by investigating its corresponding Lagrange relaxation problem ($\L(\lambda_1,\lambda_2)$). Before we give the main results, we define the following set,
\begin{align}
\X:=\big\{ Y\in \L^2_{\F_T}(\Omega;\R)~|~\gamma\leq Y\leq B, \E[Y]\geq d, \E[z(T)Y]=x_0\big\}.\label{def_X}
\end{align}
Considering the convexity issue of function $\E\big[(\gamma-x(T))_+^q  \big]$, we separate the cases with $q\geq 1$ from the ones with $0\leq q < 1$ in problem $(\A^q)$. The optimal terminal wealth of problem ($\A^q$) is given by the following two theorems separately for these two situations.

%******************************thm_lpm_q>1******************************************
\begin{thm}\label{thm_lpm_q>1}
When $q>1$, the optimal solution of problem ($\A^q$) takes one of the following two forms. (i) If $\mathcal{X}=\emptyset$, the optimal solution can be expressed as
  \begin{align}
    x^*(T)=B\1_{\eta z(T) \leq \lambda} +\left(\gamma-\left(\frac{\eta z(T)-\lambda}{q}\right)^{\frac{1}{q-1}}  \right) \1_{\lambda< \eta z(T)\leq \lambda+q\gamma^{q-1}}, \label{thm_lpm_q>1_X}
\end{align}
where the Lagrange multipliers $\eta> 0$ and $\lambda \geq 0$ satisfy the following conditions,
\begin{align}
&B\E\left [\1_{\eta z(T)\leq \lambda } \right]
+ \E\left[\left( \gamma-\left(\frac{\eta z(T)-\lambda}{q}\right)^{\frac{1}{q-1}} \right) \1_{\lambda<\eta z(T)\leq \lambda+q \gamma^{q-1}}\right] \geq d, \label{thm_lpm_q>1_eq1}\\
&B\E\left [z(T)\1_{\eta z(T) \leq \lambda }\right]
+\E\left[z(T)\left(\gamma-\left(\frac{\eta z(T)-\lambda}{q}\right)^{\frac{1}{q-1}}\right) \1_{ \lambda< \eta z(T)\leq \lambda+q\gamma^{q-1}}\right]=x_0, \label{thm_lpm_q>1_eq2}
\end{align}
and if inequality (\ref{thm_lpm_q>1_eq1}) holds strictly, $\lambda=0$.
(ii) If $\mathcal{X}\not=\emptyset$, then any random variable $x^*(T)\in \X$ is optimal for problem $(\A^q)$.

\end{thm}

\proof To simplify the notation, we use $z$ and $X$ for $z(T)$ and $x(T)$, respectively, in the following discussion. Introducing Lagrange multipliers $\lambda \geq 0$ and $\eta \in \R$, respectively, for constraints (\ref{A_lpm_cnst1}) and (\ref{A_lpm_cnst2}) in problem ($\A^q$) yields the following Lagrange relaxation of problem $(\A^q)$,
\begin{align}
\min_{0\leq X \leq B}~\hat{g}(X) = \E\Big[ (\gamma-X)^q_+ -\lambda(X-d) +\eta(zX-x_0) \Big]. \label{lpm_g0}
\end{align}
Ignoring the constant terms, we solve first the inner point-wise optimization problem,
\begin{align}
\min_{0\leq X \leq B} g(X)=(\gamma-X)^q_+ - \lambda X +\eta zX. \label{lpm_g}
\end{align}
We first assume that $\eta>0$. Since $z$ is a random variable, the optimal solution of problem (\ref{lpm_g}) depends on different values of $z$. If $\gamma-X\geq 0$, problem (\ref{lpm_g}) reduces to
\begin{align}
  \min_{0\leq X \leq \gamma } g(X)=(\gamma-X)^q-\lambda X+\eta z X. \label{lpm_g(i)}
\end{align}
It can be verified that $g(X)$ is convex with respect to $X$ in the range $0\leq X \leq \gamma$. The stationary point of function $g(X)$ satisfies
\begin{align*}
\nabla g(X)=-q(\gamma-X)^{q-1}+\eta z-\lambda=0,~~\Rightarrow~~\hat{X}=\gamma-(\frac{\eta z-\lambda}{q})^{\frac{1}{q-1}}.
\end{align*}
If $0\leq \eta z -\lambda \leq q\gamma^{q-1}$, then we have $0\leq \hat{X}\leq \gamma$. Thus, the optimal solution of problem (\ref{lpm_g(i)}) is $X^*=\hat{X}$ with $g(X^*)=(1-q)(\frac{\eta z -\lambda}{q})^{\frac{q}{q-1}}+(\eta z-\lambda)\gamma$. If $\eta z-\lambda \leq 0$, then $g(X)$ is a monotonically decreasing function with respect to $X$, which implies that the optimal solution of problem (\ref{lpm_g(i)}) is $X^*=\gamma$ with $\hat{g}(X^*)=\gamma(\eta z-\lambda)$. If $\eta z-\lambda \geq q \gamma^{q-1}$, the optimal solution is $X^*=0$ with $g(X^*)=\gamma^q$. If $\gamma - X \leq 0$, problem (\ref{lpm_g}) becomes
  \begin{align}
  \min_{\gamma \leq X \leq B} g(X)=-\lambda X+\eta zX.
  \end{align}
The optimal solution is $X^*=B$ with $g(X^*)=(\eta z-\lambda)B$ if $\eta z -\lambda <0$ and $X^*=\gamma$ with $g(X^*)=(\eta z-\lambda)\gamma$ if $\eta z-\lambda \geq 0$. As a summary, we can conclude that, when $\eta>0$, the optimal solution of problem (\ref{lpm_g0}) is
\begin{align}
X^*=\begin{dcases}
          B, ~& ~\hbox{if}~~ \eta z \leq \lambda; \\
          \gamma-\left(\frac{\eta z-\lambda}{q}\right)^{\frac{1}{q-1}}, ~&~ \hbox{if}~~ \lambda < \eta z \leq \lambda +q\gamma^{q-1}; \\
          0, ~& ~\hbox{if}~~ \lambda +q\gamma^{q-1} < \eta z.
        \end{dcases}\label{thm_lpm_q>1_X2}
\end{align}
Due to Lemma \ref{lem_lag}, we know that the Lagrange method provides the necessary and sufficient condition of problem $(\A^q)$. Thus, $X^*$ solves problem ($\A^q$) when it satisfies the conditions given in (\ref{thm_lpm_q>1_eq1}) and (\ref{thm_lpm_q>1_eq2}) for case (i).

If $\eta<0$ or $\eta=0$ and $\lambda>0$, we know that $g(X)$ is a strictly decreasing function with respect $X$. Thus, the optimal solution of problem (\ref{lpm_g0}) is $X^*=B$, which never satisfies the constraint (\ref{A_lpm_cnst2}). The only remaining case is $\eta=0$ and $\lambda=0$. Under this situation, the relaxation problem (\ref{lpm_g}) degenerates to $\min_{0\leq X\leq B}g(x)=(\gamma-X)_+^q$. That is to say, any $X^*$ that satisfies $\gamma\leq X^* \leq B$ is the optimal solution of problem (\ref{lpm_g0}). Thus, $X^*$ is the solution of problem $(\A^q)$ when $X^*$ also satisfies (\ref{A_lpm_cnst1}) and (\ref{A_lpm_cnst2}). Equivalently, we can use the set $\X$ as in (\ref{def_X}) to characterize such a solution. We thus complete the proof for case (ii).
\endproof

%*************************** thm_lpm_q<1 ***********************************
\begin{thm}\label{thm_lpm_q<1}
When $0\leq q \leq 1$, the following holds true for problem ($\A^q$):
(i) If $X=\emptyset$, the following solution solves problem problem ($\A^q$),
\begin{align}
x^*(T)=(B-\gamma)\1_{\eta z(T)\leq \lambda } +\gamma \1_{\eta z(T)\leq \lambda+\gamma^{q-1}},\label{thm_lpm_q<1_X}
\end{align}
where $\eta>0$ and $\lambda\geq 0$ satisfy the following conditions,
\begin{align}
&B\E[\1_{\eta z(T)\leq \lambda}]+\gamma \E[\1_{\lambda<\eta z(T)\leq \lambda+\gamma^{q-1}}]\geq d, \label{thm_lpm_q<1_eq1}\\
&B\E[z(T)\1_{\eta z(T)\leq \lambda}]+\gamma\E[z(T)\1_{\lambda<\eta z(T)\leq  \lambda+\gamma^{q-1} }] =x_0. \label{thm_lpm_q<1_eq2}
\end{align}
In addition, if inequality (\ref{thm_lpm_q<1_eq1}) holds strictly, then $\lambda=0$.
(ii) If $\X\not =\emptyset$ where $\X$ is defined in (\ref{def_X}), then any random variable $x^*(T)\in \X$ is optimal to problem $(\A^q)$ . (iii) When $q=1$, if problem ($\A^q$) has optimal solution,  the solution can only take one of the forms in case (i) or (ii).
\end{thm}
\proof We use the notations similar to the ones in the proof of Theorem \ref{thm_lpm_q>1}. Fisrt we assume $\eta>0$. When $0\leq X \leq\gamma$, problem (\ref{lpm_g(i)}) is to minimize a concave function with respect to $X$. Thus, the minimizer of (\ref{lpm_g(i)}) can only be at the boundary points,  either $X^*=0$ with $g(X)=\gamma^q$ or $X^*=\gamma$ with $g(X)=-\lambda \gamma+\eta z \gamma$, respectively. Comparing the function values of the two boundary points yields $X^*=0 $ if $ \eta z>\gamma^{q-1}+\lambda$ and $X^*=\gamma$ if $\eta z\leq \gamma^{q-1}+\lambda$. Combined with the case where $\gamma-X\leq 0$ in problem (\ref{lpm_g}), we have the solution of the Lagrange relaxation problem (\ref{lpm_g0}) as follows,
  \begin{align*}
  X^*=\begin{dcases}
          B, & ~~\hbox{if} ~\eta z \leq \lambda; \\
          \gamma, & ~~\hbox{if}~ \lambda < \eta z \leq \lambda +\gamma^{q-1}; \\
          0, &~~ \hbox{if} ~\lambda +\gamma^{q-1} < \eta z.
        \end{dcases}
  \end{align*}
From Lemma \ref{lem_lag}, we know that $X^*$ is the optimal solution of problem $(\A^q)$ if $X^*$ satisfies (\ref{thm_lpm_q<1_eq1}) and (\ref{thm_lpm_q<1_eq2}) given in case (i). The proof of case (ii) is the same as the proof of case (ii) in Theorem \ref{thm_lpm_q>1}, which is correspondent to the situation $\eta=0$ and $\lambda=0$. When $q=1$, the objective function of problem $(\A^{q})$ is convex. From Lemma \ref{lem_lag}, the Lagrange method characterizes all the solutions of $(\A^q)$.
\endproof

From Theorems \ref{thm_lpm_q>1} and \ref{thm_lpm_q<1}, we know that the optimal terminal wealth  $x^*(T)$ of problem $(\P_{lpm}^q)$ depends on different values of Lagrange multipliers $\lambda$ and $\eta$, which in turn depend on the parameters of problem ($\P_{lpm}^q$), e.g., the target $d$ and benchmark $\gamma$. We will discuss in details the relationship between these parameters and the Lagrange multipliers in Section \ref{sse_Lagrange}.

%modified end

%************************* the above part are modified *************************
%********************************************************************************
Theoretically speaking, once the optimal terminal wealth $X^*$ is known, the optimal portfolio policy can be characterized by solving the following backward stochastic differential equation(BSDE),
\begin{align}
\begin{dcases}
  dx(t)=\big(r(t)x(t)+ \theta(t)^{\prime}u(t) \big)dt+u(t)^{\prime}dW(t), \\
  x(T) =X^*.
\end{dcases}\label{BSDE}
\end{align}
Let $x^*(t)$ and $u^*(t)$ be the solution of the linear BSDE in (\ref{BSDE}). Then the optimal portfolio policy $\pi^*(t)$ satisfies $u^*(t)=\sigma(t)^{\prime}\pi^*(t)$. In a general setting, there could be no explicit solution for the BSDE in (\ref{BSDE}). However, when the market parameters are deterministic, the optimal wealth process and portfolio policy can be expressed explicitly, as will be shown in Section \ref{sse_determin}. From the BSDE in (\ref{BSDE}) we can get the upper and lower bounds of the whole wealth process. Recall that a no-bankruptcy constraint at the terminal time, $x(T)\geq 0$, actually ensures no-bankruptcy for the entire wealth process, i.e., $x(t)\geq 0$, for $t\in [0,T]$ (see \cite{Bielecki:2005}). Similarly, we will show now that the upper bound $B$ imposed on the terminal wealth  also induces an upper bound on the entire wealth process.

\begin{prop}\label{prop_bound}
In problem ($\P_{lpm}^q$), if we have $U\leq x(T)\leq B$, where $0\leq U <B$, then the wealth process is bounded by
\begin{align}
U\E\left[\frac{z(T)}{z(t)}~|~\F_t \right]\leq x^*(t) \leq B \E\left[\frac{z(T)}{z(t)}~|~\F_t\right],~ a.s., \label{prop_bound_eq1}
\end{align}
where process $z(t)$ is defined in (\ref{def_z}).
\end{prop}
\proof Let us consider the following BSDE with boundary condition of $x(T)=B$,
\begin{align}
\begin{dcases}
  dx(t)=\big(r(t)x(t)+ \theta(t)^{\prime}u(t) \big)dt+u(t)^{\prime}dW(t), \\
  x(T) =B.
\end{dcases}\label{prop_bound_bsde}
\end{align}
According to \cite{Karoui:1997}, the solution of (\ref{prop_bound_bsde}) can be expressed as
\begin{align}
\bar{x}(t)=B \E[\frac{z(T)}{z(t)} | \F_t].
\end{align}
By using the comparison theorem (Theorem 2.2 in \cite{Karoui:1997}), we can conclude that $x^*(t)\leq \bar{x}(t)$ for $t\in [0,T], a.s.$. We can use the similar argument for the lower bound of $x^*(t)$.
\endproof

%*******************************************************************************

%*******************************************************************************
\subsection{The existence of Lagrange multipliers}\label{sse_Lagrange}
From Theorems \ref{thm_lpm_q>1} and \ref{thm_lpm_q<1}, we know that the Lagrange multipliers $\lambda $ and $\eta$ for problem ($\A^q$) can be determined by checking the conditions in (\ref{thm_lpm_q>1_eq1}) and (\ref{thm_lpm_q>1_eq2}) for $q>1$; or the conditions in (\ref{thm_lpm_q<1_eq1}) and (\ref{thm_lpm_q<1_eq2}) for $0\leq q \leq 1$, respectively. However, at this point, we cannot guarantee the existence and uniqueness of these Lagrange multiplies. Furthermore, it is unclear at this stage under which condition the optimal solution of problem ($A^q$) takes the form (i) or (ii) in both Theorems \ref{thm_lpm_q>1} and \ref{thm_lpm_q<1}. As pointed out in \cite{Jin:2008}, the existence of the Lagrange multipliers is related to the well-poseness of the problem itself. On the other hand, from the application viewpoint, investors often adjust the investment target $d$ to generate different efficient portfolios for comparison. Thus, it is important to investigate the impact of the parameters $d$ and $\gamma$ on problem $(\A^q)$. Before we state the main result, we need the following assumption.

\begin{asmp}\label{asmp_pdf}
The probability density function of $z(T)$, $\psi(\cdot)$, is a continuous function.
\end{asmp}

We first define, for some positive number $p>0$, some functions of $p$-th order partial moments with respect to random variable $z(T)$,
\begin{align*}
H_p(y)  &:=\E[(z(T))^p\1_{z(T)\leq y}],\\
K_p(y)  &:=H_1(y)-H_{p+1}(y)/y^p,\\
J_p(y)  &:=H_0(y)-H_{p}(y)/y^p.
\end{align*}
 Obviously, when $p=0$, $H_0(y)$ reduces to the distribution function of $z(T)$. From the definition of $z(T)$, it is evident that $H_p(y)$ is a monotonically increasing function with respect to $y$, for $y>0$. Under Assumption \ref{asmp_pdf}, we can also show that $K_p(y)$ and $J_p(y)$ are also monotonically increasing functions for $y>0$. Essentially, taking the derivative of $K_p(y)$ and $J_p(y)$  with respect to $y$ gives rise to
\begin{align*}
\frac{d K_p(y) }{d y}&=\frac{dH_1(y)}{dy}-\Big(\frac{dH_{p+1}(y)/dy}{y^p}-\frac{pH_{p+1}(y)}{y^{p+1}}\Big)
=\frac{pH_{p+1}(y)}{y^{p+1}}>0,\\
\frac{d J_p(y) }{d y}&=\frac{dH_0(y)}{dy}-\Big(\frac{dH_p(y)/dy}{y^p}-\frac{pH_p(y)}{y^{p+1}}\Big)
=\frac{pH_p(y)}{y^{p+1}}>0,
\end{align*}
which imply the monotonicity of $K_p(y)$ and $J_p(y)$ for $y>0$. %The monotonicity enables us to derive the bound of these functions, e.g.,
%\begin{align*}
%&\sup_y H_0(y)=\lim_{y\rightarrow +\infty}H_0(y)=1, ~\inf_y H_0(y)=\lim_{y\rightarrow 0}H_0(y)=0,\\
%&\sup_y H_1(y)=\lim_{y\rightarrow +\infty}H_1(y)=\E[z(T)],~\inf_y H_1(y)=\lim_{y\rightarrow 0}H_1(y)=0.
%\end{align*}

For a given problem ($\P_{lpm}^q$), we define the following parameters $\underline{d}$ and $\bar{d}$, which play key roles in solving problem $(\P_{lpm}^q)$,
\begin{align}
\underline{d} &:=
\begin{dcases}
\gamma J_{p}\left(K_{p}^{-1}\left(\frac{x_0}{\gamma}\right)\right) ~\hbox{with}~p=\frac{1}{q-1}, ~~~\textrm{if}~ x_0 < \gamma \E[z(T)],~ q>1,\\
\gamma H_0\left(H_1^{-1}\left(\frac{x_0}{\gamma}\right)\right),~~ ~\textrm{if}~x_0 <\gamma \E[z(T)],~ 0\leq q\leq 1,  \\
(B-\gamma)H_0\left(H_1^{-1}\left(\frac{x_0-\gamma \E[z(T)]}{B-\gamma}\right)\right)
+\gamma, ~~~\textrm{if} ~x_0 \geq \gamma \E[z(T)],
\end{dcases}\label{def_underd}\\
\bar{d} &:=B H_0\left(H_1^{-1}\left(x_0/B\right)\right).\label{def_bard}
\end{align}
Note that $K_{p}^{-1}\left(x_0/\gamma\right)$ and $H_1^{-1}\left(x_0/B\right)$ are well defined. Due to the monotonicity of $K_p(\cdot)$, letting $y\rightarrow 0$ and $y\rightarrow +\infty$ yields $\inf_{y>0}  K_p(y)=0$ and $\sup_{y>0} K_p(y)=\E[z(T)]$, respectively. The condition $x_0<\gamma \E[z(T)]$ implies the existence of $K_{p}^{-1}\left(x_0/\gamma\right)$. The similar argument also applies to $H^{-1}_1(x_0/B)$ and $H^{-1}_1(x_0/\gamma)$.

The following propositions ensure the existence and uniqueness of the Lagrange multipliers $\lambda$ and $\eta$ in Theorems \ref{thm_lpm_q>1} and \ref{thm_lpm_q<1}.
%***************************************************************************
\begin{prop}\label{prop_lagq>1}
For problem $(\A^q)$ with $q>1$, under Assumption \ref{asmp_pdf}, the following results hold.
\begin{itemize}
  \item [(i)] If $\underline{d}<  d < \bar{d}$, the solution of problem ($\A^q$) is given by (\ref{thm_lpm_q>1_X}) and there is a unique pair of $\lambda>0$ and $\eta>0$ satisfying the conditions in (\ref{thm_lpm_q>1_eq1}) and (\ref{thm_lpm_q>1_eq2}) with equality holding for (\ref{thm_lpm_q>1_eq1}).

  \item [(ii)] If $d \leq \underline{d}$ and $x_0 <\gamma \E[z(T)] $, the solution of problem ($\A^q$) is given by (\ref{thm_lpm_q>1_X}) with  $\lambda=0$ and $\eta=q \gamma^{q-1}/K^{-1}_{\frac{1}{q-1}}(x_0/\gamma)$ satisfying the conditions in (\ref{thm_lpm_q>1_eq1}) and (\ref{thm_lpm_q>1_eq2}).

  \item [(iii)] If $d \leq \underline{d}$ and $x_0 \geq \gamma \E[z(T)] $, then problem $(\A)$ has multiple solutions which are characterized by set $\X$ defined in (\ref{def_X}) and one of such solutions is
      \begin{align}
      x^*(T)=(B-\gamma)\1_{z(T)\leq H_1^{-1}(\frac{x_0-\gamma\E[z(T)]}{B-\gamma})}+\gamma.\label{degen_X}
      \end{align}
\end{itemize}
\end{prop}

\proof (i) We prove this result by identifying the range of $d$ under which the equality holds for both (\ref{thm_lpm_q>1_eq1}) and (\ref{thm_lpm_q>1_eq2}). To simplify our notation, we change the variables $\lambda $ and $\eta$ in conditions (\ref{thm_lpm_q>1_eq1}) and (\ref{thm_lpm_q>1_eq2}) to
\begin{align}
\delta:= \lambda/\eta, ~~\rho := q \gamma ^{q-1}/\eta, \label{prop_lagq>1_map}
\end{align}
respectively. Note that $0\leq \delta$ and $0< \rho $. In the following part, we let $p=1/(q-1)$. When both equalities hold, the conditions in (\ref{thm_lpm_q>1_eq1}) and (\ref{thm_lpm_q>1_eq2}) become a system of two equations,
\begin{align}
&I_1(\delta,\rho)=d, \label{prop_lagq>1_eq1}\\
&I_2(\delta,\rho)=x_0,\label{prop_lagq>1_eq2}
\end{align}
where
\begin{align*}
&I_1(\delta,\rho):= BH_0(\delta)+\gamma\Big(H_0(\delta+\rho)-H_0(\delta) \Big)  -\frac{\gamma}{\rho ^p} \int _{\delta}^{\delta +\rho}(s-\delta)^p\psi(s)ds, \\
&I_2(\delta,\rho):= BH_1(\delta)+\gamma\Big(H_1(\delta+\rho)-H_1(\delta) \Big)  -\frac{\gamma}{\rho ^p} \int _{\delta}^{\delta +\rho}s(s-\delta)^p\psi(s)ds.
%%(1+\frac{\delta}{\rho})
\end{align*}
We show that $I_1(\delta,\rho)$ and $I_2(\delta,\rho)$ are monotonically increasing functions with respect to both $\delta$ and $\rho$. To simplify the notations, we do not write out the arguments in $I_1(\delta, \rho)$ and $I_2(\delta,\rho)$ explicitly, unless necessary. Assumption \ref{asmp_pdf} implies the differentiability of $I_1$ and $I_2$. Taking the derivatives of $I_1$ and $I_2$ with respect to $\delta$  and $\rho$, respectively, yields the following results for $\delta>0$ and $\rho>0$,
\begin{align}
\begin{dcases}
\frac{\partial I_1}{\partial \delta}=(B-\gamma) \psi(\delta)
+\frac{p \gamma}{\rho ^p} \int _{\delta}^{\delta +\rho}(s-\delta)^{p-1}\psi(s)ds>0,\\
\frac{\partial I_1}{\partial \rho}=\frac{p \gamma}{\rho ^{p+1}} \int _{\delta}^{\delta +\rho}(s-\delta)^{p}\psi(s)ds>0,\\
\frac{\partial I_2}{\partial \delta}=(B-\gamma)\delta \psi(\delta)
+\frac{p \gamma}{\rho ^p} \int _{\delta}^{\delta +\rho}s(s-\delta)^{p-1}\psi(s)ds>0,\\
\frac{\partial I_2}{\partial \rho}=\frac{p \gamma}{\rho ^{p+1}} \int _{\delta}^{\delta +\rho}s(s-\delta)^{p}\psi(s)ds>0,
\end{dcases}\label{prop_lagq>1_eq0}
\end{align}
which imply the monotonicities of the $I_1$ and $I_2$ with respect to both $\delta$ and $\rho$. Thus, we can determine the ranges of $I_1$ and $I_2$ as $0< I_1(\delta,\rho) < I_1(\infty,\infty)=B$ and $0< I_2(\delta,\rho)< I_2(\infty,\infty)=B\E[z(T)]$, respectively. To solve the system of equations (\ref{prop_lagq>1_eq1}) and (\ref{prop_lagq>1_eq2}), we define the following function $I_3(\rho): \mathbb{R}_+\rightarrow \mathbb{R} $ as follows,
\begin{align*}
I_3(\rho):=  I_1(\hat{\delta},\rho)~\textrm{with $\hat{\delta}$ satisfying $I_2(\hat{\delta},\rho)=x_0$}.
\end{align*}
For a given $\rho>0$, due to the monotonicity of $I_2$, $0$ $\leq I_2(\eta,\rho)$ $\leq$ $I_2(\infty,\rho)= B\E[z(T)]$ holds true. Thus, there exists a unique $\hat{\delta}$ that satisfies $I_2(\hat{\delta},\rho)=x_0$. Thus, $I_3(\rho)$ is well defined. We will show that $I_3(\rho)$ is a monotonically decreasing function with respect to $\rho$. Since $I_2(\delta,\rho)=x_0$  holds, we have
\begin{align}
0=\frac{\partial I_2}{\partial \delta}d \delta+\frac{\partial I_2}{\partial \rho}d\rho, ~~\Rightarrow~~\frac{d \delta}{d \rho}=-\left(\frac{\partial I_2}{\partial \rho}\right)/\left(\frac{\partial I_2}{\partial \delta}\right). \label{prop_lagq>1_delta_rho}
\end{align}
Checking the derivative of $I_3(\rho)$ with respect to $\rho$ gives rise to
\begin{align}
\frac{d I_3(\rho)}{d \rho}&=\frac{\partial I_1}{\partial \delta}\frac{d \delta}{d \rho}+\frac{\partial I_1}{\partial \rho},\notag \\
&=\left(-\frac{\partial I_1}{\partial \delta} \frac{\partial I_2 }{\partial \rho}+\frac{\partial I_1}{\partial \rho} \frac{\partial I_2}{\partial \delta}    \right)/(\frac{\partial I_2}{\partial \delta} ), \label{prop_lagq>1_dI3}
\end{align}
where the last equality is based on (\ref{prop_lagq>1_delta_rho}). Since $\partial I_2/\partial \delta>0$, the sign of $d I_3(\rho)/ d \rho$ depends on the numerator of (\ref{prop_lagq>1_dI3}). Combining (\ref{prop_lagq>1_eq0}) with (\ref{prop_lagq>1_dI3}) further yields the following,
\begin{align}
\frac{\partial I_1}{\partial \rho} \frac{\partial I_2}{\partial \delta}-\frac{\partial I_1}{\partial \delta} \frac{\partial I_2 }{\partial \rho}
&=\frac{(p\gamma)^2}{\rho^{2p+1}}\Bigg[\Big(\int_{\delta}^{\delta+\rho}s(s-\delta)^{p-1}\psi(s)ds \Big)^2\notag\\
&~~~~~-\int_{\delta}^{\delta+\rho}s^2(s-\delta)^{p-1}\psi(s)ds\cdot \int_{\delta}^{\delta+\rho}(s-\delta)^{p-1}\psi(s)ds\Bigg]\notag\\
&~~~~~-\frac{p \gamma (B-\gamma) \psi(\delta)}{\rho^{p+1}}\Big( \int_{\delta}^{\delta+\rho} (s-\delta)^{p+1}\psi(s)ds \Big)<0,
 \label{prop_lagq>1_eq3}
\end{align}
where the last inequality is from the Cauchy Schwarz Inequality,
\begin{align*}
& \left(\int _{\delta}^{\delta +\rho}s(s-\delta)^{p-1}\psi(s)ds\right)^2 \\
&~~~~~~<  \int _{\delta}^{\delta +\rho}\Big(s(s-\delta)^{\frac{p-1}{2}}\psi(s)^{\frac{1}{2}}\Big)^2ds
\cdot\int _{\delta}^{\delta +\rho}\Big((s-\delta)^{\frac{p-1}{2}}\psi(s)^{\frac{1}{2}}\Big)^2ds.
\end{align*}
 Inequality (\ref{prop_lagq>1_eq3}) implies that $d I_3(\rho)/d \rho<0$, which proves the monotonicity of $I_3(\rho)$. Thus, for any $d$, if $d$ is in the range space of $I_3(\rho)$, we can always find a unique $\rho$ such that $I_3(\rho)=d$. Now, we only have to fix the range of $I_3(\rho)$. Due to the definition of $I_3(\rho)$, we define
\begin{align}
\underline{\rho}&:=\inf\{\rho \in \R~|~I_2(\delta, \rho)=x_0, \delta \geq 0, \rho>0 \},\label{prop_lagq>1_underrho}\\
\bar{\rho}&:=\sup\{\rho \in \R~|~I_2(\delta, \rho)=x_0, \delta \geq 0, \rho>0 \}. \label{prop_lagq>1_barrho}
\end{align}
From (\ref{prop_lagq>1_eq0}) and (\ref{prop_lagq>1_delta_rho}), we know that $d \rho /d \delta<0$, if $I_2(\delta,\rho)=x_0$ holds. It is not hard to see $\underline{\rho}=0$. We can find the corresponding $\delta$ when $\rho \rightarrow 0$ as follows,
\begin{align*}
\lim_{\rho \rightarrow 0} I_2(\delta,\rho)= BH_1(\delta)=x_0,~~\Rightarrow~~\bar{\delta}:=H_1^{-1}(x_0/B).
\end{align*}
Taking $\rho\rightarrow \underline{\rho}$ and $\delta\rightarrow \bar{\delta}$ yields the upper limit of $I_3(\rho)$,
\begin{align}
\sup_{\rho>0} I_3(\rho)=\lim_{\delta \rightarrow \bar{\delta}, \rho \rightarrow 0 }I_1(\delta,\rho)=BH_0(H_1^{-1}(x_0/B)). \label{prop_lagq>1_I3_ub}
\end{align}
Now, we focus on the lower limit of $I_3(\rho)$. Since $d\rho/d \delta<0$ when $I_2(\delta,\rho)=x_0$ holds, there are two candidates of $\bar{\rho}$, namely, $\infty$ or the corresponding $\rho$ when $\delta\rightarrow 0$. If $x_0\leq \gamma\E[z(T)]$, we find $\bar{\rho}$ by checking
\begin{align*}
\lim_{\delta\rightarrow 0} I_2(\delta, \bar{\rho})=\gamma H_1(\bar{\rho})-\gamma H_{p+1}(\bar{\rho})=\gamma K_p(\bar{\rho})=x_0,~~\Rightarrow~~  \bar{\rho}=K_p^{-1}(x_0/\gamma),
\end{align*}
which leads to the lower limit of $I_{3}(\rho)$,
\begin{align}
\inf_{0< \rho \leq \bar{\rho}}I_3(\rho) =\lim_{\delta\rightarrow 0,\rho\rightarrow\bar{\rho}}I_1(\delta,\rho)=\gamma J_p(K_p^{-1}(x_0/\gamma)).\label{prop_lagq>1_I3_lb1}
\end{align}
If $x_0\geq \gamma\E[z(T)]$, we have $\bar{\rho}=\infty$. We can identify the corresponding $\underline{\delta}$ in $I_2(\delta,\rho)=x_0$ as
\begin{align}
(B-\gamma)H_1(\delta)+\gamma\E[z(T)]=x_0, ~~\Rightarrow~~ \underline{\delta}=H_1^{-1}(\frac{x_0-\gamma\E[z(T)]}{B-\gamma}), \label{prop_lagq>1_eq4}
\end{align}
which further gives rise to
\begin{align}
\inf_{\rho>0}I_3(\rho)=\lim_{\rho\rightarrow \infty, \delta\rightarrow \underline{\delta}}=(B-\gamma)H_0\left( H_1^{-1}\left( \frac{x_0-\gamma\E[z(T)]}{B-\gamma}\right)\right)+\gamma.\label{prop_lagq>1_I3_lb2}
\end{align}
As a summary, the upper and lower limits of $I_3(\rho)$ in (\ref{prop_lagq>1_I3_ub}) and (\ref{prop_lagq>1_I3_lb1}) are just $\bar{d}$ and $\underline{d}$ defined in (\ref{def_bard}) and (\ref{def_underd}), respectively. Since $I_3(\rho)$ is a monotonically decreasing function with respect to $\rho$, if $d\in (\underline{d}, \bar{d})$,  we can find a unique $\rho^*>0$ that solves $I_3(\rho^*)=d$. Due to the monotonicity of $I_2(\delta,\rho)$, we can further substitute $\rho^*$ back to $I_2(\delta,\rho)$ to solve for $\delta^*>0$. Note that the pairs $(\lambda, \eta)$ and $(\delta, \rho)$ are of a one-to-one mapping from (\ref{prop_lagq>1_map}), which completes the proof of (i) in this proposition.

(ii) We first consider the case when $d=\underline{d}$. From (\ref{prop_lagq>1_I3_lb1}) and $x_0 < \gamma \E[z(T)]$, we know that both (\ref{prop_lagq>1_eq1}) and (\ref{prop_lagq>1_eq2}) hold, when $\delta=0$ and $\rho=\bar{\rho}=K_p^{-1}(x_0/\gamma)$, which further implies $\lambda=0$ and $\eta=q\gamma^{q-1}/K_p^{-1}(x_0/\gamma)$ by (\ref{prop_lagq>1_map}). When $d< \underline{d}$, we can easily verify that $\lambda$ and $\eta$ also satisfy the conditions in (\ref{thm_lpm_q>1_eq1}) and (\ref{thm_lpm_q>1_eq2}) in Theorem \ref{thm_lpm_q>1}.

(iii) When $d=\underline{d}$ and $x_0\geq \gamma\E[z(T)]$, from (\ref{prop_lagq>1_I3_lb2}), we know that both (\ref{prop_lagq>1_eq1}) and (\ref{prop_lagq>1_eq2}) hold for $\delta^*=H_1^{-1}(\frac{x_0-\gamma\E[z(T)]}{B-\gamma})$ and $\rho\rightarrow \infty$ which implies $\lambda= 0$ and $\eta=0$. From the proof for (ii) of Theorem \ref{thm_lpm_q>1}, we know that any $x^*(T)$ in set $\X$ is optimal to problem $(\A^q)$. It is not hard to verify $x^*(T)$ given in (\ref{degen_X}) to be one of such solutions.
\endproof

%*****************************************************************************
\begin{prop}\label{prop_lagq<1}
For problem $(\A^{q})$ with $0\leq q \leq 1$, the following results hold.
\begin{itemize}
  \item [(i)] If $\underline{d}< d < \bar{d}$, the solution given in (\ref{thm_lpm_q<1_X}) solves problem ($\A^q$) and there is a unique pairs of $\lambda>0$ and $\eta>0$ satisfying the conditions in (\ref{thm_lpm_q<1_eq1}) and (\ref{thm_lpm_q<1_eq2}) with equality holding for (\ref{thm_lpm_q<1_eq1}).

  \item [(ii)] If $ d \leq \underline{d}$ and $x_0 <\gamma \E[z(T)] $, the solution given in (\ref{thm_lpm_q<1_X}) solves problem ($\A^q$) with $\lambda=0$ and $\eta=\gamma^{q-1}/H^{-1}_1(x_0/\gamma)$ satisfying the conditions in (\ref{thm_lpm_q>1_eq1}) and (\ref{thm_lpm_q>1_eq2}).

  \item [(iii)] If $d \leq \underline{d}$ and $x_0 \geq \gamma \E[z(T)] $, the set $\X$ defined in (\ref{def_X}) is nonempty and any $X\in\X$ is a solution of problem $(\A^q)$. One of such solutions can be constructed as in (\ref{degen_X}).
\end{itemize}
\end{prop}

\proof (i) Similar to the proof of Proposition \ref{prop_lagq>1}, we identify the range of $d$ under which the equalities hold for both conditions (\ref{thm_lpm_q<1_eq1}) and (\ref{thm_lpm_q<1_eq2}). We change the variables as
\begin{align}
\delta := \lambda/\eta,~~~\rho:=\gamma^{q-1}/\eta.\label{prop_lagq<1_map}
\end{align}
Clearly, we have $0\leq \delta $ and $0< \rho $. When the equalities hold for both conditions in (\ref{thm_lpm_q<1_eq1}) and (\ref{thm_lpm_q<1_eq2}), we have the following system of two equations,
\begin{align}
I_1(\delta,\rho)&=d, \label{prop_lagq<1_eq1}\\
I_2(\delta,\rho)&=x_0, \label{prop_lagq<1_eq2}
\end{align}
where $I_1(\delta, \rho)$ and $I_2(\delta,\rho)$ are defined as
\begin{align*}
I_1(\delta,\rho)&:=BH_0(\delta)+\gamma \big(H_0(\delta+\rho)-H_0(\delta)\big), \\
I_2(\delta,\rho)&:=BH_1(\delta)+\gamma \big(H_1(\delta+\rho)-H_1(\delta)\big).
\end{align*}
For any $\delta_1>\delta_2>0$, we have
\begin{align*}
I_1(\delta_1,\rho)-I_2(\delta_2,\rho)=
(B-\gamma)(H_0(\delta_1)-H_0(\delta_2))
+\gamma(H_0(\rho+\delta_1)-H_0(\rho+\delta_2))>0,
\end{align*}
which implies $I_1(\delta,\rho)$ is monotonically increasing with respect to $\delta$. Using the similar procedure, we can prove that both $I_1(\delta,\rho)$ and $I_2(\delta,\rho)$ are monotonically increasing with respect to $\delta$ and $\rho$, respectively. Rearranging $(\ref{prop_lagq<1_eq2})$ gives rise to

%To solve the system of equations (\ref{prop_lagq<1_eq1}) and (\ref{prop_lagq<1_eq2}), we eliminate $\rho$ by representing $\delta +\rho$ in equation $(\ref{prop_lagq<1_eq2})$, e.g.,
\begin{align}
H_1(\delta+\rho)=\frac{x_0-(B-\gamma)H_1(\delta)}{\gamma}.
\label{prop_lagq<1_deltarho}
\end{align}
The expression in (\ref{prop_lagq<1_deltarho}) helps us obtain both upper and lower limits of $\delta$, i.e.,
\begin{align*}
\bar{\delta}&:=\sup \{\delta \in \R ~|~ I_2(\delta, \rho)=x_0,\delta\geq 0, \rho>0 \}, \\
\underline{\delta}&:=\inf\{ \delta \in \R~|~I_2(\delta,\rho)=x_0, \delta\geq 0, \rho>0\}.
\end{align*}
Due to the monotonicity of $H_1(\cdot)$, we have $0\leq H_1(\delta)< H_1(\delta+\rho)< H_1(\infty)=\E[z(T)]$, which further leads to
\begin{align}
H_1(\delta)< \frac{x_0-(B-\gamma)H_1(\delta)}{\gamma}<\E[z(T)],~~\Rightarrow~
~\frac{x_0-\gamma \E[z(T)]}{B-\gamma}<H_1(\delta)\leq \frac{x_0}{B}. \label{prop_lagq<1_ineq4}
\end{align}
The inequality in (\ref{prop_lagq<1_ineq4}) provides the lower and  upper limits of $\delta$ as follows,
\begin{align}
\underline{\delta}&=
\begin{dcases}
0 &\hbox{if} ~~x_0< \gamma \E[z(T)],\\
H_1^{-1}\left(\frac{x_0-\gamma \E[z(T)]}{B-\gamma}\right) &\hbox{if}~~x_0\geq \gamma \E[z(T)],
\end{dcases}\label{prop_lagq<1_ld}\\
\bar{\delta}&=H_1^{-1}(x_0/B). \label{prop_lagq<1_ud}
\end{align}
From (\ref{prop_lagq<1_deltarho}), since $\delta \in (\underline{\delta},\bar{\delta})$, we also have
\begin{align}
\delta+\rho=H_1^{-1}\left(\frac{x_0-(B-\gamma)H_1(\delta)}{\gamma}\right).
\label{prop_lagq<1_delta}
\end{align}
Substituting $\delta+\rho$ in (\ref{prop_lagq<1_delta}) back to (\ref{prop_lagq<1_eq1}) yields the following function $L(\delta)$,
\begin{align}
L(\delta):= (B-\gamma)H_0(\delta)+\gamma H_0\left(H_1^{-1}\left(\frac{x_0-(B-\gamma)H_1(\delta)}{\gamma} \right)\right). \label{prop_lagq<1_L}
\end{align}
We first prove that $L(\delta)$ is a monotonically increasing function with respect to $\delta$. Considering $\delta_1$ and $\delta_2$ satisfying $\underline{\delta}<\delta_1<\bar{\delta}$ and $\underline{\delta}<\delta_2<\bar{\delta}$ with $\delta_1>\delta_2$, we have
\begin{align*}
L(\delta_1)&:= (B-\gamma)H_0(\delta_1)+\gamma H_0\left(H_1^{-1}\left(\frac{x_0-(B-\gamma)H_1(\delta_1)}{\gamma}\right)\right),\\
L(\delta_2)&:= (B-\gamma)H_0(\delta_2)+\gamma H_0\left(H_1^{-1}\left(\frac{x_0-(B-\gamma)H_1(\delta_2)}{\gamma}\right)\right).
\end{align*}
Let $\rho_1$ and $\rho_2$ satisfy
\begin{align}
\delta_1+\rho_1&=H_1^{-1}\left(\frac{x_0-(B-\gamma)H_1(\delta_1)}{\gamma}\right),~~
\Rightarrow~
(B-\gamma)H_1(\delta_1)+\gamma H_1(\delta_1+\rho_1)=x_0 ,\label{zeta_1}\\
\delta_2+\rho_2&=H_1^{-1}\left(\frac{x_0-(B-\gamma)H_1(\delta_2)}{\gamma}\right),~~
\Rightarrow~
(B-\gamma)H_1(\delta_2)+\gamma H_1(\delta_2+\rho_2)=x_0\label{zeta_2}.
\end{align}
From (\ref{prop_lagq<1_delta}) and the monotonicity of $H_1(\cdot)$, we have
$\delta_2+\rho_2> \delta_1 +\rho_1$. The difference between (\ref{zeta_1}) and (\ref{zeta_2}) gives rise to
\begin{align}
\gamma \E[z(T)\1_{\delta_1+\rho_1\leq z(T) \leq \delta_2+\rho_2}]=(B-\gamma)\E[z(T)\1_{\delta_2 \leq z(T) \leq \delta_1 } ]. \label{prop_lagq<1_diff}
\end{align}
The following inequalities then hold,
\begin{align}
&\gamma(\delta_1+\rho_1) \E[\1_{\delta_1+\rho_1\leq z(T) \leq \delta_2+\rho_2}]\leq \gamma \E[z(T)\1_{\delta_1+\rho_1\leq z(T) \leq \delta_2+\rho_2}],\label{prop_lagq<1_ineq1}\\
&(B-\gamma)\E[z(T)\1_{\delta_2 \leq z(T) \leq \delta_1 } ]\leq \delta_1(B-\gamma)\E[\1_{\delta_2\leq z(T) \leq \delta_1}].\label{prop_lagq<1_ineq2}
\end{align}
Combining (\ref{prop_lagq<1_ineq1}), (\ref{prop_lagq<1_ineq2}) and (\ref{prop_lagq<1_diff}) yields
\begin{align}
\gamma(\delta_1+\rho_1) \E[\1_{\delta_1+\rho_1\leq z(T)\leq \delta_2+\rho_2}]\leq \delta_1(B-\gamma)\E[\1_{\delta_2\leq z(T)\leq \delta_1}]. \label{prop_lagq<1_ineq3}
\end{align}
The inequality in (\ref{prop_lagq<1_ineq3}) implies,
\begin{align*}
L(\delta_1)-L(\delta_2)&=(B-\gamma )\E[\1_{\delta_1\leq z(T) \leq \delta_2}]-\gamma\E[\1_{\delta_1+\rho_1\leq z(T)\leq \delta_2+\rho_2}] \nonumber\\
&\geq (B-\gamma)\left(\frac{\rho_1}{\delta_1+\rho_1}\right)\E[\1_{\delta_2\leq z(T) \leq \delta_1}]>0.
\end{align*}
Thus, function $L(\delta)$ is monotonically increasing with respect to $\delta$. Then, we can identify the range of $L(\delta)$ as,
\begin{align}
&\inf_{\delta} L(\delta)=\lim_{\delta \rightarrow \underline{\delta}} L(\delta)\label{prop_lagq<1_infL} \\
&~~~~~~~=\begin{dcases}
\gamma H_0\left(  H_1^{-1}(x_0/\gamma)\right) & \hbox{if}~~x_0<\gamma \E[z(T)],\\
(B-\gamma)H_0\left( H_1^{-1}\left( \frac{x_0-\gamma \E[z(T)]}{B-\gamma}\right)\right)+\gamma & \hbox{if} ~~x_0\geq \gamma \E[z(T)],
\end{dcases}\notag\\
&\sup_{\delta} L(\delta)= \lim_{\delta \rightarrow \bar{\delta}} L(\delta)
=BH_0\left(  H_1^{-1}(x_0/B)\right), \label{prop_lagq<1_supL}
\end{align}
where (\ref{prop_lagq<1_infL}) and (\ref{prop_lagq<1_supL}) are exactly
 the constants $\underline{d}$ and $\bar{d}$ defined in (\ref{def_underd}) and (\ref{def_bard}), respectively. Thus, if $d\in(\underline{d},\bar{d})$, due to the monotonicity of $L(\delta)$, we can find a unique $\underline{\delta}<\delta^*<\bar{\delta}$ such that $L(\delta^*)=d$. Again, by the monotonicity, the unique solution $\rho^*>0$ can be found by substituting $\delta^*$ into (\ref{prop_lagq<1_deltarho}). Note that the two pairs of $(\delta^*, \rho^*)$ and $(\lambda, \eta)$ are of a one-to-one mapping from (\ref{prop_lagq<1_map}), which completes the proof of case (i).

(ii) We first consider the case $d=\underline{d}$ and $x_0< \gamma \E[z(T)]$. From (\ref{prop_lagq<1_ud}) and (\ref{prop_lagq<1_infL}), we know that the system of two equations in (\ref{prop_lagq<1_eq1}) and (\ref{prop_lagq<1_eq2}) hold when $\delta=\underline{\delta}=0$ with the correspondent $\rho=\hat{\rho}= H^{-1}_1(x_0/\gamma)$ by (\ref{prop_lagq<1_deltarho}). Due to the one-to-one mapping in (\ref{prop_lagq<1_map}), we have $\lambda=0$ and $\eta^*=\gamma^{q-1}/H_1^{-1}(x_0/\gamma)$. When $d<\underline{d}$, it can be verified that $\lambda$ and $\eta$ satisfy both conditions in (\ref{thm_lpm_q<1_eq1}) and (\ref{thm_lpm_q<1_eq2}) with strictly inequality holding for (\ref{thm_lpm_q<1_eq1}).

(iii) Similar to case (iii) in Proposition \ref{prop_lagq>1}, when $d=\underline{d}$ and $x_0\geq \gamma \E[z(T)]$, from (\ref{prop_lagq<1_infL}), the system of two equations in (\ref{prop_lagq<1_eq1}) and (\ref{prop_lagq<1_eq2}) has the solution of $\delta=\underline{\delta}=H_1^{-1}(
\frac{x_0-\gamma\E[z(T)]}{B-\gamma})$ and $\rho\rightarrow \infty$, which further implies that $\eta=0$ and $\lambda=0$. From the proof for item (ii) of Theorem \ref{thm_lpm_q<1}, we have the result in (ii).
\endproof

%*********************************************************************************
Note that Assumption \ref{asmp_pdf} is necessary for the proof of Proposition \ref{prop_lagq>1}, since we need to use the differentiability of $H_p(y)$. However, for Proposition \ref{prop_lagq<1}, such an assumption is not needed.

Propositions \ref{prop_lagq>1} and \ref{prop_lagq<1} reveal the relationship between the parameters ($d$ and $\gamma$) and the existence and uniqueness of the Lagrange multiplies for problem ($\P_{lpm}^q$). Only when $d\in (\underline{d}, \bar{d})$, the two Lagrange multiplies in problem ($A^q$) are both positive, or in other words, the two constraints in (\ref{A_lpm_cnst1}) and (\ref{A_lpm_cnst2}) are truly active in the problem. We classify this case as the \textbf{ regular cases}. When $d<\underline{d}$, the parameter $d$ does not affect the problem any more. We classify this situation as the \textbf{degenerated cases}. As a summary, we list in Table \ref{Table_x} the conditions for different situations under which the optimal terminal wealth $x^*(T)$ of problem $(\A^q)$ is determined.

\begin{table}[h] \label{Table_x}
  \centering
  \begin{tabular}{lll}
    \toprule
    Condition     &  $q>1$     &  $0\leq q\leq1$ \\
                  &  $x^*(T)$ is determined by & $x^*(T)$ is determined by\\
    \hline
    $\underline{d}<d<\bar{d}$ &  case (i) in Proposition \ref{prop_lagq>1}    & case (i) in Proposition \ref{prop_lagq<1} \\
    $d\leq \underline{d}$, $x_0 < \gamma \E[z(T)]$     & case (ii) in Proposition \ref{prop_lagq>1}  &  case (ii) in Proposition \ref{prop_lagq<1} \\
    $d\leq \underline{d}$, $x_0 \geq \gamma \E[z(T)]$  & case (iii) in Proposition \ref{prop_lagq>1} & case (iii) in Proposition \ref{prop_lagq<1} \\
    \bottomrule
  \end{tabular}
  \caption{Classification of the optimal terminal wealth $x^*(T)$ of problem $(\A^q)$}
\end{table}

%The degenerated case is also related to the following problem.
%\begin{align*}
%(\tilde{\P}_{lpm}^{q})~~&~~\min_{\pi(\cdot)\in \L^2_{\F}(0,T; \R^n) }~~\E[(\gamma-x(T) )_+^q ]\\
%\textrm{Subject to}~&~\begin{dcases}
%                        \textrm{($x(\cdot)$,$\pi(\cdot)$) statisfies dynamics (\ref{def_wealth}) }, \\
%                        0 \leq x(T) \leq B,
%                        %\pi(\cdot) \in \L_{\F}^2(0,T;\R^n),
%                      \end{dcases}%\label{def_P1_constraint}
%\end{align*}
%which has a clear financial meaning, it generates the global minimum $q$-th order LPM. It is just the counter part problem to the global variance minimization in well-known mean-variance analysis. We do not list the solution of problem ($\tilde{\P}^q_{lpm}$) explicitly, since the  solution of such problem is just as same as the case $d\leq \underline{d}_q$ or $d\leq \underline{d}$ in Proposition \ref{prop_lagq>1}, Proposition \ref{prop_lagq<1} and Proposition \ref{prop_degen}.

In problem ($\P_{lpm}^q$), especially for the regular cases, it will be interesting to investigate the probability of attaining upper bound $B$. Denote the probability that the optimal terminal wealth $x^*(t)$ reaches upper bound $B$ as  $\mP(x^*(T)=B)$. Then, the following result is true.

\begin{prop}\label{prop_B}
In problem ($\P_{lpm}^q$), if $\underline{d}<d<\bar{d}$, then the probability $\mP(x^*(T)=B)$ is monotonically decreasing with respect to $B$. Furthermore, if $x_0<\gamma \E[z(T)]$, we have
\begin{align*}
\lim_{B\rightarrow \infty}\mP(x^*(T)=B) =0.
\end{align*}
\end{prop}

\proof (i) We first prove the result for $q>1$. Similar to Proposition \ref{prop_lagq>1}, we replace the variables $\lambda$ and $\eta$ by $\delta=\lambda/\eta$ and $\rho=(\lambda+\gamma)/\eta$, respectively, in (\ref{thm_lpm_q>1_eq1}) and (\ref{thm_lpm_q>1_eq2}), which leads to (\ref{prop_lagq>1_eq1}) and (\ref{prop_lagq>1_eq2}). Since Assumption \ref{asmp_pdf} holds, we can check the following total differential for both (\ref{prop_lagq>1_eq1}) and (\ref{prop_lagq>1_eq2}),
\begin{align*}
& \frac{\partial I_1}{\partial \delta} d \delta+\frac{\partial I_1}{\partial \rho}d \rho + \frac{\pp I_1}{\pp B}dB=0,\\
&\frac{\partial I_2}{\partial \delta} d \delta+\frac{\partial I_2}{\partial \rho}d \rho + \frac{\pp I_2}{\pp B}dB=0.
\end{align*}
Solving these two equations by eliminating $d\rho$ yields the following,
\begin{align}
\frac{d\delta}{d B}=\frac{\frac{\partial I_1}{\partial \rho}H_1(\delta)-\frac{\partial I_2}{\partial \rho}H_0(\delta)}{
\frac{\partial I_1}{\partial \delta}\frac{\partial I_2}{\partial \rho}-\frac{\partial I_1}{\partial \rho}\frac{\partial I_2}{\partial \delta}}.\label{prop_B_eq1}
\end{align}
From (\ref{prop_lagq>1_eq3}), we know that the denominator of (\ref{prop_B_eq1}) is positive, i.e.,
\begin{align*}
\frac{\partial I_1}{\partial \delta}\frac{\partial I_2}{\partial \rho} - \frac{\partial I_1 }{\partial \rho}\frac{\partial I_2}{\partial \delta}>0.
\end{align*}
We now check the sign of the numerator of (\ref{prop_B_eq1}). By using (\ref{prop_lagq>1_eq0}), we have
\begin{align*}
&\frac{\partial I_1}{\partial \rho}H_1(\delta)-\frac{\partial I_2}{\partial \rho}H_0(\delta)\notag \\
&=\frac{\gamma p}{\rho^{p+1}} \left[ \int^{\delta+\rho}_{\delta} (s- \delta)^{p} \psi(s)ds \int^{\delta}_{-\infty} \tau\psi(\tau)d\tau - \int^{\delta+\rho}_{\delta} s(s- \delta )^p\psi(s)ds\int^{\delta}_{-\infty} \psi(\tau)d\tau  \right] \notag \\
&= \frac{\gamma p}{\rho ^{p+1}} \int^{\delta+\rho}_{\delta}\int^{\delta}_{-\infty}  (\tau - s)(s - \delta  )^p \psi(\tau)\psi(s)ds d\tau <0.
%\label{prop_lagq>1_equ5}
\end{align*}
Thus, we can conclude that $d \delta/dB<0$, which further implies that $\delta$ is decreasing when $B$ increases. Note that the probability $\mP(x^*(T)=B)=\mP(z(T)\leq \delta)=H_0(\delta)$ is a monotonically increasing function of $\delta$. Thus, the probability  $\mP(x^*(T)=B)$ is decreasing with respect to $B$. From (\ref{def_underd}) and (\ref{def_bard}), if $x_0< \gamma \E[z(T)]$, we know $\underline{d}$ is irrelative to $B$ and $\bar{d}$ is increasing when $B$ increases. Thus, for a given $d\in (\underline{d},\bar{d})$, $d$ will remain in the interval $(\underline{d},\bar{d})$ when $B$ increases. From (\ref{prop_lagq>1_eq1}), we have $I_1(\delta,0)\leq I_1(\delta,\rho)=d$, which implies $H_0(\delta)\leq d/B$. Thus, when $B\rightarrow \infty$, $H_0(\delta)$ is monotonically decreasing to $0$.

(ii) We use the similar notations to these in Proposition \ref{prop_lagq<1} to prove the case with $0 \leq q \leq 1$. Suppose that $\delta$ and $\rho$ solve both (\ref{prop_lagq<1_eq1}) and (\ref{prop_lagq<1_eq2}). We show now that $\delta$ monotonically decreases when $B$ increases. Particularly, let $B_1>B_2$ and $\delta_1$, $\rho_1$, $\delta_2$ and $\rho_2$ solve the following two systems of two equations in (\ref{prop_lagq<1_eq1}) and (\ref{prop_lagq<1_eq2}), i.e.,
\begin{align}
       \begin{dcases}
       B_1H_0(\delta_1)+\gamma \big(H_0(\delta_1+\rho_1)-H_0(\delta_1)\big)=d,\\
       B_1H_1(\delta_1)+\gamma \big(H_1(\delta_1+\rho_1)-H_1(\delta_1)\big)=x_0,
       \end{dcases}\label{prop_Bq<1_eq1}\\
       \begin{dcases}
       B_2H_0(\delta_2)+\gamma \big(H_0(\delta_2+\rho_2)-H_0(\delta_2)\big)=d,\\
       B_2H_1(\delta_2)+\gamma \big(H_1(\delta_2+\rho_2)-H_1(\delta_2)\big)=x_0.
       \end{dcases} \label{prop_Bq<1_eq2}
\end{align}
We would like to prove $\delta_2>\delta_1$. From the definitions of $\delta$ and $\rho$ in (\ref{prop_lagq<1_map}), we have $\delta_1<\delta_1+\rho_1$ and $\delta_2<\delta_2+\rho_2$. Note that if $\delta_2>\delta_1+\rho_1$, then we have $\delta_2>\delta_1$, which completes our proof for the monotonicity. Thus, we only need to consider the case of $\delta_2<\delta_1+\rho_1$. For any $y_1<y_2<\delta_1+\rho_1$, we have
\begin{align}
&\left(H_0(y_2)-\frac{H_1(y_2)}{\delta_1+\rho_1}\right)-
\left(H_0(y_1)-\frac{H_1(y_1)}{\delta_1+\rho_1}\right)\notag \\
&=\E[\1_{y_1\leq z(T)\leq y_2}]-\frac{1}{\delta_1+\rho_1}\E[z(T)\1_{y_1\leq z(T)\leq y_2}]\notag\\
&\geq (1-\frac{y_2}{\delta_1+\rho_1})\E[z(T)\1_{y_1\leq z(T)\leq y_2}]>0, \label{prop_Bq<1_FG}
\end{align}
which further implies that $(H_0(y)-\frac{1}{\delta_1+\rho_1}H_1(y))$ is monotonically increasing with respect to $y$ for $y>0$. Checking the difference between the first equations in (\ref{prop_Bq<1_eq1}) and (\ref{prop_Bq<1_eq2}) gives rise to
\begin{align}
(B_1-\gamma)H_0(\delta_1)- (B_2-\gamma)H_0(\delta_2) & =\gamma\big(H_0(\delta_2+\rho_2)-H_0(\delta_1+\rho_1)\big). \label{prop_Bq<1_dif1}
\end{align}
If $\delta_2+\rho_2<\delta_1+\rho_1$, due to the monotonicity of $H_0(\cdot)$, we have
\begin{align*}
(B_1-\gamma)H_0(\delta_1)<(B_2-\gamma)H_0(\delta_2),
\end{align*}
which further implies $\delta_2>\delta_1$. Next we consider the case of $\delta_2+\rho_2>\delta_1+\rho_1$, under which (\ref{prop_Bq<1_dif1}) becomes
\begin{align}
(B_1-\gamma)H_0(\delta_1)-(B_2-\gamma)H_0(\delta_2)
&=\gamma(\E[\1_{\delta_1+\rho_1\leq z(T) \leq \delta_2+\rho_2}]). \label{prop_Bq<1_dif11}
\end{align}
Similarly, checking the difference of the second equations in (\ref{prop_Bq<1_eq1}) and (\ref{prop_Bq<1_eq2}) yields,
\begin{align}
(B_1-\gamma)H_1(\delta_1)-(B_2-\gamma)H_1(\delta_2)
&= \gamma \E[z(T)\1_{\delta_1+\rho_1\leq z(T)\leq \delta_2+\rho_2}]\notag \\
&~~~\geq \gamma (\delta_1+\rho_1)\E[\1_{\delta_1+\rho_1\leq z(T)\leq \delta_2+\rho_2}]. \label{prop_Bq<1_dif2}
\end{align}
Combining (\ref{prop_Bq<1_dif11}) with (\ref{prop_Bq<1_dif2}) gives rise to
\begin{align*}
(B_1-\gamma)\left(H_0(\delta_1)-\frac{H_1(\delta_1)}{\delta_1+\rho_1}\right)\leq (B_2-\gamma)\left(H_0(\delta_2)-\frac{H_1(\delta_2)}{\delta_1+\rho_1}\right).
\end{align*}
We already show in (\ref{prop_Bq<1_FG}) that $H_0(y)-\frac{1}{\delta_1+\rho_1}H_1(y)$ is a monotonically increasing function of $y$ for $y<\delta_1+\rho_1$. Thus, $\delta_1<\delta_2$ and we can conclude that $\delta$ is monotonically decreasing when $B$ increases, which further implies $\mP(x^*(T)=B)=H_0(\delta)$ is a monotonically decreasing function of $B$. When $x_0<\gamma \E[z(T)]$, from (\ref{def_underd}) and (\ref{def_bard}), we know that $d\in (\underline{d},\bar{d})$ while we increase $B$. As the probability $\mP(x^*(T)=B)\geq 0$ and $H_0(\delta)<d/B$, when $B$ goes to infinity, we have $\lim_{B\rightarrow \infty} \mP(x^*(T)=B)=0 $.
\endproof
%************************************************************

\subsection{Special market setting with a deterministic opportunity set}\label{sse_determin}
In this section, we consider the case where the market parameters are deterministic, i.e., the following assumption holds.
\begin{asmp}\label{assumption_deterministic}
The risk free return rate $r(t)$, the drift rate $\mu_i(t)$, $i=1,\cdots, n$, and volatility $\sigma_{ij}(t)$, $i,j=1,\cdots,n$, are all deterministic functions of $t$ for $t\in[0,T]$.
\end{asmp}

Under Assumption \ref{assumption_deterministic}, we can derive the explicit expressions for the optimal wealth process and the portfolio process for problem ($\P_{lpm}^q$). Note that the definition of the deflator process $z(t)$ in (\ref{def_z}) implies that $z(T)/z(t)$ follows a log-normal distribution. In other words, $\ln\big(z(T)/z(t)\big)$ follows a normal distribution with its mean  $m(t)$ and variance $\nu^2(t)$ given as
\begin{align}
m(t)&  = -\int_{t}^T (r(s)+\frac{1}{2}\|\theta(s)\|^2 )d \tau, ~~~t\in[0,T],\label{def_m(t)}\\
\nu^2(t)& = \int_{t}^T \| \theta(s) \|^2 ds, ~~~t\in[0,T].\label{def_v(t)}
\end{align}
As a special case, when $t=0$,  $\ln(z(T))$ follows the normal distribution with mean and variance being $m(0)$ and $\nu^2(0)$, respectively. Furthermore, we can also compute $\E[z(T)]=e^{-\int_{0}^T r(s)ds}$.

Under Assumption \ref{assumption_deterministic}, Proposition \ref{prop_bound} reduces to
 \begin{align*}
U e^{-\int_{t}^T r(s)ds}\leq x^*(t) \leq B e^{-\int_{t}^T r(s)ds},
\end{align*}
as we can compute $\E[z(T)/z(t)|\F_t]$ explicitly by using Lemma \ref{lem_expectation} in the Appendix.

We first investigate how the parameters $\underline{d}$ and $\bar{d}$ in (\ref{def_underd}) and (\ref{def_bard}) become more explicit in such a market setting with a deterministic opportunity set.

\begin{prop}\label{prop_d}
Under Assumption \ref{assumption_deterministic}, we have the following for problem $(\P_{lpm}^q)$,
\begin{align}
\label{prop_d_hatd}
\underline{d} &=\begin{dcases}
\gamma \Big(\Phi\big(F(\bar{\rho})\big) -e^{-\int_0^T r(s)ds} \Phi\big( F(\bar{\rho})\\
~~~~~~~~-v(0)\big)/\bar{\rho}  \Big) & \hbox{if}~x_0<\gamma e^{-\int_0^Tr(s)ds}, ~q=2,\\
\gamma \Phi\Big( F(\hat{\rho})\Big) & \hbox{if}~x_0<\gamma e^{-\int_0^Tr(s)ds},~0\leq q \leq 1,\\
(B-\gamma)\Phi\big(F(\underline{\delta})\big)+\gamma &\hbox{if}~x_0\geq \gamma e^{-\int_0^Tr(s)ds},
\end{dcases}\\
\label{prop_d_bard}\bar{d}  &=B\Phi(F(\bar{\delta})),
\end{align}
where $\bar{\rho}$, $\underline{\delta}$, $\hat{\rho}$ and $\bar{\delta}$ are the solutions to the following four equations, respectively,
\begin{align}
&e^{-\int^{T}_0r(s)ds}\Phi\Big(F(\bar{\rho})-\nu(0)\Big)
+\frac{e^{2m(0)+2\nu^2(0)}}{\bar{\rho}}\Phi\Big( F(\bar{\rho})-2\nu(0)\Big)=\frac{x_0}{\gamma}, \label{prop_d_barrho}\\
&e^{-\int^{T}_0r(s)ds}\Phi\Big(F(\underline{\delta})-\nu(0) \Big)=\frac{x_0-\gamma e^{-\int^T_0r(s)ds}}{B-\gamma}, \label{prop_d_underdelta}\\
&e^{-\int^{T}_0r(s)ds}\Phi\Big(F(\hat{\rho})-\nu(0)\Big)=\frac{x_0}{\gamma}, \label{prop_d_hatrho}\\
&e^{-\int^{T}_0r(s)ds}\Phi\Big(F(\bar{\delta})-\nu(0) \Big)=\frac{x_0}{B},\label{def_bardelta}
\end{align}
with $F(y):=(\ln(y)-m(0))/\nu(0)$.
\end{prop}

\proof Let us consider first the case with $x_0<\gamma e^{-\int_0^T r(s)ds}$ and $q=2$.  From the definition of $\underline{d}$ in (\ref{def_underd}), we have
\begin{align*}
K_1(\bar{\rho})&=\E[z(T)\1_{z(T)\leq \bar{\rho}}]-\E[z^2(T)\1_{z(T)\leq \bar{\rho}}]/\bar{\rho}\\
&=\E[e^{\ln(z(T))}\1_{\ln(z(T))\leq \ln(\bar{\rho})} ]-\E[e^{2\ln(z(T))}\1_{\ln(z(T))\leq \ln(\bar{\rho})} ]/\bar{\rho}=x_0/\gamma,
\end{align*}
which leads to (\ref{prop_d_barrho}) by using Lemma \ref{lem_expectation}.  Then we can compute $\underline{d}$ as
\begin{align*}
\underline{d}&=\gamma\big(H_0(\bar{\rho})-H_1(\bar{\rho})/\bar{\rho}\big)\\
&=\gamma \Big(\E[\1_{\ln(z(T))\leq \ln(\bar{\rho})}]-\E[e^{\ln(z(T))}\1_{\ln(z(T))\leq \ln(\bar{\rho})}]/\bar{\rho} \Big),
\end{align*}
which gives the first case in (\ref{prop_d_hatd}).  We can compute $\underline{d}$ and $\bar{d}$ for other cases in similar ways.
\endproof

The following two theorems offer the explicit optimal wealth process and optimal portfolio policy of problem $({\P}_{lpm}^{q})$ for cases with $q=2$ and $0\leq q\leq 1$, respectively.

\begin{thm}\label{thm_lpm_dq>1}
Under Assumption \ref{assumption_deterministic}, the optimal solution of problem  (${\P}_{lpm}^{2}$) is given as follows.
(i) If $\underline{d}<d<\bar{d}$, the optimal wealth process is
  \begin{align}
  x^*(t)&=e^{m(t)+\frac{\nu^2(t)}{2}}
  \Big((B-\gamma-\frac{\lambda}{2})\Phi\big(K_1(t)-\nu(t)\big)
  +(\gamma+\frac{\lambda}{2} )\Phi\big(K_2(t)-\nu(t)\big) \Big)\notag\\
  &+\frac{z(t)\eta}{2} e^{2m(t)+2\nu^2(t)}\Big(\Phi\big(K_1(t)-2\nu(t)\big)-\Phi\big(K_2(t)-2\nu(t)\big) \Big)\label{thm_lpm_dq>1_x}
  \end{align}
  and the optimal portfolio policy is
  \begin{align}
  \pi^*(t)&=\Bigg\{\frac{1}{\sqrt{2\pi}\nu(t)}e^{m(t)+\frac{\nu^2(t)}{2}}
  \Big[ (B-\gamma-\frac{\lambda}{2})e^{-\frac{(K_1(t)-\nu(t))^2}{2}}+(\gamma +\frac{\lambda}{2})e^{-\frac{(K_2(t)-\nu(t))^2}{2}}\Big]\notag\\
   &~~-\frac{\eta z(t)}{2} e^{2m(t)+2\nu^2(t) }\Big[\Phi\big(K_1(t)-2\nu(t)\big)-\Phi\big(K_2(t)-2\nu(t)\big)\notag \\
   &~~-\frac{1}{\sqrt{2\pi}\nu(t)}\big(e^{-\frac{(K_1(t)-2\nu(t))^2}{2}} -e^{-\frac{(K_2(t)-2\nu(t) )^2}{2}}\big) \Big]\Bigg \} \big(\sigma(t)\sigma(t)^{\prime}\big)^{-1}b(t), \label{thm_lpm_dq>1_pi}
  \end{align}
   where $K_1(t)$ and $K_2(t)$ are defined by
   \begin{align*}
    K_1(t)=\frac{\ln\left(\frac{\lambda}{\eta z(t)}\right)-m(t)}{\nu(t)} ,~~K_2(t)=\frac{ \ln\left(\frac{\lambda+2\gamma}{\eta z(t)}\right)-m(t) }{\nu(t)},
   \end{align*}
and $\lambda >0$ and $\eta >0$ are the unique solutions of the following two equations,
\begin{align}
&\label{thm_lpm_dq>1_eq1}(B-\gamma-\frac{\lambda}{2})\Phi\left(K_1(0)\right)+(\gamma+\frac{\lambda}{2})\Phi(K_2(0))\\
&~~~~~~~+\frac{\eta}{2}e^{m(0)+\frac{\nu^2(0)}{2}}
\left(\Phi\left(K_1(0)-\nu(0)\right)  -\Phi\left(K_2(0)-\nu(0)\right)\right)=d,\notag\\
&\label{thm_lpm_dq>1_eq2}e^{m(0)+\frac{\nu^2(0)}{2}}\left((B-\gamma-\frac{\lambda}{2})\Phi(K_1(0)-\nu(0))
+(\gamma+\frac{\lambda}{2})\Phi(K_2(0)-\nu(0)) \right) \\
&~~~~~~~+\frac{\eta}{2} e^{2m(0)+2\nu^2(0)}\big(\Phi(K_1(0)-2\nu(0))-\Phi(K_2(0)-2\nu(0))  \big)=x_0.\notag
\end{align}
(ii) If $d \leq \underline{d}$ and $x_0<\gamma e^{-\int_0^T r(s)ds}$, the optimal wealth process and portfolio policy are given as in (\ref{thm_lpm_dq>1_x}) and (\ref{thm_lpm_dq>1_pi}), respectively, where $\lambda=0$ and $\eta=2\gamma/\bar{\rho}$ with $\bar{\rho}$ being given in (\ref{prop_d_barrho}).\\
(iii) If $d \leq \underline{d}$ and $x_0\geq \gamma e^{-\int_0^T r(s)ds}$, the optimal wealth process and portfolio are given as
\begin{align}
x^*(t)&=e^{m(t)+\frac{\nu^2(t)}{2}}\Big((B-\gamma)\Phi\big(K_3(t)-\nu(t)\big)+\gamma \Big),\label{thm_lpm_deg_xt}\\
\pi^*(t)&=\frac{1}{\sqrt{2\pi}\nu(t)}e^{m(t)+\frac{\nu^2(t)}{2}}
(B-\gamma)e^{-\frac{(K_3(t)-2\nu(t))^2}{2}}
\big(\sigma(t)\sigma(t)^{\prime}\big)^{-1}b(t),\label{thm_lpm_deg_pi}
\end{align}
where $K_3(t):=\big(\ln\big(\underline{\delta}/z(t)\big)-m(t)\big)/\nu(t)$ with $\underline{\delta}$ being the solution of (\ref{prop_d_underdelta}).

\end{thm}

\proof (i) We first consider the case with $\underline{d}<d<\bar{d}$. The following result is true for any pair of parameters $a>0$ and $c>0$. Note that $z(t)$ is $\F_t$-adapted and $\lambda>0$ and $\eta>0$ from Proposition \ref{prop_lagq>1}. We then have
\begin{align}
&\E\left[\frac{z(T)}{z(t)}\Big(a+\frac{\eta}{2} z(T)\Big)\1_{\eta z(T)\leq \lambda +c}  ~\Big| ~\F_t\right]\notag \\
&=a\E\left[ \frac{z(T)}{z(t)}\1_{z(T)\leq \frac{\lambda+c}{\eta}}~\Big|~\F_t\right]+\frac{\eta z(t)}{2}\E\left[ (\frac{z(T)}{z(t)})^{2}\1_{z(T)\leq \frac{\lambda+c}{\eta}} ~\Big|~\F_t\right] \notag\\
&=a\E\left[e^{\ln \frac{z(T)}{z(t)}}\1_{\ln \frac{z(T)}{z(t)}\leq \ln \frac{\lambda+c}{\eta z(t)} } ~\Big|~\F_t\right]+\frac{\eta z(t)}{2} \E\left[e^{2\ln\frac{z(T)}{z(t)}} \1_{\ln \frac{z(T)}{z(t)} \leq \ln \frac{\lambda+c}{\eta z(t)} }  ~\Big|~\F_t\right]\notag\\
&=a e^{m(t) + \frac{\nu^2(t)}{2}}\Phi\left( \frac{\ln\frac{\lambda+c}{\eta z(t)}  -m(t)}{\nu(t)} -\nu(t)\right)\notag \\
&~~~~~~~~~~~+\frac{\eta z(t)}{2} e^{2m(t)+2\nu^2(t)}\Phi\left(  \frac{\ln\frac{\lambda+c}{\eta z(t)}  -m(t)}{\nu(t)} -2\nu(t)\right), \label{thm_lpm_dq>1_Ex1}
\end{align}
where the last equality is based on Lemma \ref{lem_expectation}. The discounted optimal wealth process is a martingale under the probability measure $\tilde{\mP}$ (see \cite{Karatzas:1998}), i.e., we have
\begin{align*}
x^*(t)=\E\left[\frac{z(T)}{z(t)}x^*(T)~\Big|~\F_t\right], ~t\in[0,T].
\end{align*}
From the expression of $x^*(T)$ in (\ref{thm_lpm_q>1_X}), we can compute $x^*(t)$ as
\begin{align}
x^*(t)&=\E\Big[\frac{z(T)}{z(t)}\Big( \big( B-\gamma+\frac{\eta z(T)-\lambda}{2}\big)\1_{\eta z(T)\leq \lambda}\notag\\
&~~~~~~~+ \big( \gamma-\frac{\eta z(T)-\lambda}{2} \big)\1_{\eta z(T)\leq \lambda+2\gamma} \Big)~|~\F_t\Big]. \label{thm_lpm_dq>1_Ex2}
\end{align}
Let $a=B-\gamma-\frac{\lambda}{2}$ and $c=0$ for the first part of (\ref{thm_lpm_dq>1_Ex2}) and $a=\gamma+\frac{\lambda}{2}$ and $c=2\gamma$ for the second part of (\ref{thm_lpm_dq>1_Ex2}).  Applying (\ref{thm_lpm_dq>1_Ex1}) to (\ref{thm_lpm_dq>1_Ex2}) gives rise to the result in (\ref{thm_lpm_dq>1_x}).

Under Assumption \ref{assumption_deterministic}, we can assume $x^*(t)$ as a deterministic function of $z(t)$ and $t$, i.e., there exists a function $G(\cdot,\cdot)$ such that $x^*=G(z(t),t)$. Now let us determine the functional form of $G(z(t),t)$. Applying It\"{o} Lemma yields
\begin{align}
dG(z(t),t)&=\frac{\pp G(z(t),t)}{\pp z(t)}dz(t) +\frac{\pp G(z(t),t)}{\pp t}dt +\frac{1}{2} \frac{\pp^2 G(z(t),t)}{\pp z(t)^2}(\theta(t)^2z(t)^2)dt\notag \\
&=(-z(t)\frac{\pp G(z(t),t)}{\pp z(t)}r(t)+\frac{\pp G(z(t),t)}{\pp t}+\frac{1}{2}\frac{\pp^2 G(z(t),t)}{\pp z(t)^2}z(t)^2\|\theta(t)\|^2 )dt\notag\\
&~~-\frac{\pp G(z(t),t)}{\pp z(t)}z(t)\theta(t)^{\prime} dW(t).\label{thm_lpm_dq>1_dF}
\end{align}
Comparing the diffusion term in (\ref{thm_lpm_dq>1_dF}) with the one of the wealth process in (\ref{def_wealth}) dictates the following,
\begin{align}
\pi^*(t)^{\prime}\sigma(t) =-\frac{\pp G(z(t),t)}{\pp z(t)} z(t) \theta(t)^{\prime}. \label{thm_lpm_dq>1_spi}
\end{align}
From the definition of $\theta$ in (\ref{def_theta}), multiplying $\sigma(t)$ on both sides of (\ref{thm_lpm_dq>1_spi}) gives rise to
\begin{align*}
\pi^*(t)=-\frac{G(z(t),t)}{\pp z(t)}z(t)(\sigma(t)\sigma(t)^{\prime})^{-1}b(t).
\end{align*}
Thus, differentiating (\ref{thm_lpm_dq>1_x}) with respect to $z(t)$ further gives rise to the result in (\ref{thm_lpm_dq>1_pi}).

From Theorem \ref{thm_lpm_q>1} and Proposition \ref{prop_lagq>1}, we can identify Lagrange multipliers $\eta$ and $\lambda$ by solving equations (\ref{thm_lpm_q>1_eq1}) and (\ref{thm_lpm_q>1_eq2}). Equation (\ref{thm_lpm_q>1_eq2}) can be written explicitly  by letting $t=0$ in (\ref{thm_lpm_dq>1_x}) which yields (\ref{thm_lpm_dq>1_eq2}). Based on (\ref{thm_lpm_q>1_eq1}), we have
\begin{align}
\E[x^*(T)]&=\E\Big[\big(B-\gamma -\frac{\lambda}{2}+\frac{\eta}{2}z(T)\big)\1_{\eta z(T)\leq \lambda}+\big(\gamma+\frac{\lambda}{2}-\frac{\eta}{2}z(T)\big)\1_{\eta z(T)\leq \lambda+2\gamma}    \Big]\notag \\
&=\E\Big[\big(B-\gamma-\frac{\lambda}{2}\big)\1_{\ln(z(T))\leq \ln \frac{\lambda}{\eta} }+\frac{\eta}{2}e^{\ln z(T)}\1_{\ln(z(T))\leq \ln\frac{\lambda}{\eta}}\notag \\
&~~~~~~~+ (\gamma+\frac{\lambda}{2})\1_{\ln z(T)\leq \ln(\frac{\lambda+2\gamma}{\eta})}-\frac{\eta}{2} e^{\ln z(T)}\1_{\ln z(T)\leq \ln(\frac{\lambda+2\gamma}{\eta})}    \Big].\label{thm_lpm_dq>1_ExT}
\end{align}
Applying Lemma \ref{lem_expectation} to (\ref{thm_lpm_dq>1_ExT}) yields (\ref{thm_lpm_dq>1_eq1}).
%The optimal objective value can be computed by using (\ref{prop_lpm_X_q>1}),
%\begin{align}
%\E[(\gamma-x^*(T))_+^2]=\E[(\frac{\eta z(T)-\lambda}{2})^2 \1_{\frac{\lambda}{\eta}\leq z(T)\leq \frac{\lambda+2\gamma}{\eta} }]+\gamma\E[\1_{z(T)\geq \frac{\lambda+2\gamma}{\eta} }]. \label{thm_lpm_dq>1_E}
%\end{align}
%By using Lemma \ref{lem_expectation}, we can get  (\ref{thm_lpm_dq>1_eq2}) and (\ref{thm_lpm_dq>1_obj}) from (\ref{thm_lpm_dq>1_ExT}) and (\ref{thm_lpm_dq>1_E}), respectively.

(ii) From Proposition \ref{prop_lagq>1}, we know that $x^*(T)$ takes the same form as in (\ref{thm_lpm_dq>1_x}) with the Lagrange multipliers $\lambda=0$ and $\eta=2\gamma/K_1^{-1}(x_0/\gamma)$. Note that $K_1^{-1}(x_0/\gamma)$ is nothing but $\bar{\rho}$ given in (\ref{prop_d_barrho}).

(iii) From Proposition \ref{prop_lagq>1}, there are multiple optimal solutions and one of the optimal terminal wealth is given in (\ref{degen_X}). Then $x^*(t)$ and $\pi^*(t)$ can be computed similarly as in case (i) by using Lemma \ref{lem_expectation}. From the definition of $\underline{\delta}$ in (\ref{prop_d_underdelta}), we know that $\underline{\delta}=H_1^{-1}( \frac{x_0-\gamma \E[z(T)]}{B-\gamma})$.
\endproof

%****************************** theorem p=1 ************************************
\begin{thm}\label{thm_lpm_dq<1}
Under Assumption \ref{assumption_deterministic}, the optimal solution of problem (${\P}_{lpm}^{q}$) with $0\leq q \leq 1$ is given as follows.
(i) If $\underline{d}<d <\bar{d}$, the optimal wealth process is
\begin{align}
x^*(t)&=e^{m(t)+\frac{\nu^2(t)}{2}}\bigg((B-\gamma)\Phi\big( \bar{K}_1(t)-\nu(t)\big)+
\gamma\Phi\big(\bar{K}_2(t)-\nu(t)\big) \bigg)\label{thm_lpm_dq<1_x}
\end{align}
and the optimal portfolio policy is
\begin{align}
\pi^*(t)&=\frac{e^{m(t)+\frac{\nu^2(t)}{2}}}{\sqrt{2\pi}\nu(t)} \bigg((B-\gamma)e^{-\frac{(\bar{K}_1(t)-\nu(t))^2}{2}} +\gamma e^{-\frac{(\bar{K}_2(t)-\nu(t))^2}{2}}\bigg)
(\sigma(t)\sigma(t)^{\prime})^{-1}b(t), \label{thm_lpm_dq<1_pi}
\end{align}
with the optimal objective value being
\begin{align}
E[(\gamma-x^*(T))_+^q]=\gamma^q (1-\Phi(K_2(0))), \label{thm_lpm_dq<1_obj}
\end{align}
where $\bar{K}_1(t)$ and $\bar{K}_2(t)$ are defined as
\begin{align*}
\bar{K}_1(t)=\frac{\ln\left(\frac{\lambda}{\eta z(t)}\right)-m(t)}{\nu(t)},
~\bar{K}_2(t)= \frac{\ln\left(\frac{\lambda+\gamma^{q-1}}{\eta z(t)}\right)-m(t)}{\nu(t)},
\end{align*}
with $\lambda > 0$ and $\eta>0$ being the solution of the following two equations,
\begin{align}
&(B-\gamma)\Phi\big(\bar{K}_1(0)\big)+\gamma\Phi\big(\bar{K}_2(0)\big)=d,
\label{thm_lpm_dq<1_eq1}\\
&(B-\gamma)\Phi\big(\bar{K}_1(0)-\nu(0)\big)+\gamma\Phi\big(\bar{K}_2(0)-\nu(0)\big)
=e^{-m(0)-\frac{\nu^2(0)}{2}}x_0.
\label{thm_lpm_dq<1_eq2}
\end{align}

(ii) If $d\leq \underline{d}$ and $x_0 < \gamma e^{-\int_0^T r(s)ds }$, the optimal wealth process and  portfolio policy are given as in (\ref{thm_lpm_dq<1_x}) and (\ref{thm_lpm_dq<1_pi}), respectively, where $\lambda=0$ and $\eta= \gamma^{q-1}/\hat{\rho}$ with $\hat{\rho}$ being given as in (\ref{prop_d_hatrho}).

(iii) If $d \leq \underline{d}$ and $x_0 \geq \gamma e^{-\int_0^T r(s)ds}$, the optimal wealth process and portfolio are given as in (\ref{thm_lpm_deg_xt}) and (\ref{thm_lpm_deg_pi}), respectively.
\end{thm}
\proof From  Theorem \ref{thm_lpm_q<1} and Proposition \ref{prop_lagq<1}, we can compute $x^*(t)$ and $\pi^*(t)$ by using a method similar to the proof of Theorem \ref{thm_lpm_dq>1} for cases (i), (ii) and (iii). We thus omit the details here.
\endproof

\begin{rem}\label{rem_lpm}
In Theorems \ref{thm_lpm_dq>1} and \ref{thm_lpm_dq<1},  the optimal wealth process $x^*(t)$ and optimal portfolio policy $\pi^*(t)$ are represented by the market state density $z(t)$. Although  $z(t)$ can be computed by observing the price when the market is complete, $z(t)$, in general case, cannot be observed or computed directly. Thus, it is more favorable to have the portfolio policy in a feedback form, i.e., to represent the portfolio policy $\pi^*(t)$ in terms of the current wealth $x^*(t)$. Taking the derivative of $x^*(t)$ with respect to $z(t)$ in (\ref{thm_lpm_dq>1_x}) and (\ref{thm_lpm_dq<1_x}), we can show that $x^*(t)$ is a monotonically decreasing function of $z(t)$ under a common market setting when $B$ is sufficiently large. That is to say, the expressions in (\ref{thm_lpm_dq>1_x}) and (\ref{thm_lpm_dq>1_pi}) define a one to one mapping between $x^*(t)$ and $z(t)$. Thus, theoretically, we can replace $z(t)$ by $x^*(t)$ in both (\ref{thm_lpm_dq>1_pi}) and (\ref{thm_lpm_dq<1_pi}) to achieve a feedback type of policy. Since there is no analytical form to represent $z(t)$ by $x^*(t)$ from (\ref{thm_lpm_dq>1_x}) and (\ref{thm_lpm_dq<1_x}), we should discretize $z(t)$ first and compute next the correspondent value of $\pi^*(t)$ and $x^*(t)$ for each $z(t)$. The relationship of $\pi^*(t)$ and $x^*(t)$ can be approximately achieved by using a curve fitting method.
\end{rem}

Note that the probability that $x^*(T)$ reaches the upper bound for problem $(\P_{lpm}^q)$ can be expressed as $$\mathbb{P}(x^*(T)=B)=\mathbb{P}(z(T)\leq \frac{\lambda}{\eta}).$$
When the market opportunity set is deterministic, this probability can be computed explicitly for problem ($\P_{lpm}^q$) as $\mathbb{P}(x^*(T)=B)=\Phi(\frac{\ln(\lambda/\eta)-m(0)}{\mu(0)})$,
where $\lambda$ and $\eta$ are the solution to (\ref{thm_lpm_dq>1_eq1}) and (\ref{thm_lpm_dq>1_eq2}) for problem $(\P_{lpm}^2)$ and the solution to (\ref{thm_lpm_dq<1_eq1}) and (\ref{thm_lpm_dq<1_eq2}) for problem $(\P_{lpm}^q)$ with $0\leq q\leq 1$.

\section{Optimal Portfolio Policy for Mean-CVaR Formulation}\label{se_cvar}
We solve in this section the mean-CVaR portfolio optimization model $(\P_{cvar})$. Recall the definition of the investment loss, $f(x(T))$, in (\ref{def_loss}) and the definition of CVaR of the loss in \cite{Rockafellar:2002}. As the cumulative distribution function of $f(x(T))$ is defined as
\begin{align*}
\Psi(y)=\mP(f(x(T))\leq y),
\end{align*}
the correspondent $\beta$-tail distribution for a given confidence level $\beta$ is
\begin{align}
\Psi_{\beta}(y)=\begin{dcases}
                 0, & \hbox{if}~ y<\VaR_\beta, \\
                 \frac{\Psi(y)-\beta}{1-\beta}, & \hbox{if}~ y\geq \VaR_{\beta},
               \end{dcases}
\end{align}
where $\VaR_{\beta}=\inf\{y~|~\Psi(y)\geq \beta\}$. The CVaR of the loss function $f(x(T))$ is then given as
\begin{align}
\CVaR[f(x(T))]:= \int_{f(x(T))\geq \VaR_{\beta}} f(x(T))d\Psi_{\beta}(y),\label{def_CVaR}
\end{align}
where the integration should be understood as a summation when the distribution of $y$ is discrete. Note that the above definition of CVaR is for a  general distribution function of the loss function $f(x(T))$, see for example Rockafllar and Uryasev \cite{Rockafellar:2002} for some subtle difference between the cases of discrete distributions and continuous distributions.

To solve the mean-CVaR portfolio optimization problem $(P_{cvar})$, we utilize the parameterized expression of CVaR introduced in \cite{Rockafellar:2000} and \cite{Rockafellar:2002}.
\begin{lem}\label{lem_cvar}
The CVaR of the loss $f(x(T))$ of the terminal wealth can be computed as follows,
\begin{align}
\CVaR[f(x(T))]=\min_{\alpha} \Big\{ \alpha + \frac{1}{1-\beta} \E \big[ (\bar{x}_T-x(T) -\alpha)_+ \big] \Big\}, \label{def_par_cvar}
\end{align}
where $\alpha$ is an auxiliary variable.
\end{lem}

Introducing parameter $\alpha$ and rewriting the objective function of problem $(P_{cvar})$ using (\ref{def_par_cvar}) yields the following equivalent formulation of problem (${\P}_{cvar}$),
\begin{align}
(\P_{cvar})~&~\min_{\pi(\cdot) \in \L_{\F}^2(0,T;\R^n)),~\alpha}~J(\alpha):= \alpha + \frac{1}{1-\beta} \E \big[ (\bar{x}_T-x(T) -\alpha)_+ \big] ,\label{barP_cvar_obj}\\
\textrm{Subject to}  ~&~
                      \begin{dcases}
                        \E[x(T)]\geq d, \\
                        \textrm{($x(\cdot)$,$\pi(\cdot)$) statisfies (\ref{def_wealth}) }, \\
                        0 \leq x(T)\leq B.
                      \end{dcases}
\end{align}
To solve problem $({\P}_{cvar})$, we first solve the following auxiliary problem for fixed $\alpha$,
\begin{align*}
(\P_{cvar}(\alpha)):~~&~\min_{\pi(\cdot) \in \L_{\F}^2(0,T;\R^n)} ~ \E\big[ (\bar{x}_T-\alpha-x(T))_+ \big] \\
\textrm{Subject to} ~&~ \begin{dcases}
                        \E[x(T)]\geq d, \\
                        \textrm{($x(\cdot)$,$\pi(\cdot)$) statisfies (\ref{def_wealth}) }, \\
                        0 \leq x(T)\leq B.
                      \end{dcases}
\end{align*}

The difference between problem ($\P_{cvar}$) and ($\P_{cvar}(\alpha)$) is that the decision variable $\alpha$ is fixed as a constant in problem ($\P_{cvar}(\alpha)$), which leaves $\pi(\cdot)$ as the only decision vector. Thus, we can first solve the problem ($\P_{cvar}(\alpha)$) for given $\alpha$ and then identify optimal $\alpha^*$ under which the optimal portfolio policy for ($\P_{cvar}(\alpha^*)$) also solves ($\P_{cvar}$). Once $\alpha$ is fixed in problem ($\P_{cvar}(\alpha)$), if we regard $\bar{x}_T-\alpha$ as $\gamma$, then $(\P_{cvar}(\alpha))$ takes the same form as  $(\P_{lpm}^1)$. However, when $\alpha$ varies, $\underline{d}$ is changing. From (\ref{prop_d_hatd}) for $0\leq q\leq 1$, we redefine $\underline{d}$ for a fixed $\alpha$ in problem $(\P_{cvar}(\alpha))$ as
\begin{align}
\label{def_underd_alpha}\underline{d}(\alpha)=\begin{dcases}
(\bar{x}_T-\alpha)\Phi\left( F(\hat{\rho}(\alpha))\right) & \hbox{if}~~x_0<(\bar{x}_T-\alpha)e^{-\int_0^Tr(s)ds},\\
(B-\bar{x}_T+\alpha)\Phi(F(\underline{\delta}(\alpha)))+\bar{x}_T-\alpha &\hbox{if}~~x_0\geq (\bar{x}_T-\alpha)e^{-\int_0^Tr(s)ds},
\end{dcases}
\end{align}
where $F(\cdot)$ is defined in Proposition \ref{prop_d}, and $\hat{\rho}(\alpha)$ and $\underline{\delta}(\alpha)$ are determined by the following equations,
\begin{align}
\hat{\rho}(\alpha):~&~e^{-\int^{T}_0r(s)ds}\Phi\Big(F(\hat{\rho}(\alpha))-\nu(0)\Big)
=\frac{x_0}{\bar{x}_T-\alpha}, \label{cvar_hatrho}\\
\underline{\delta}(\alpha):~&~e^{-\int^{T}_0r(s)ds}\Phi\Big(F(\underline{\delta}(\alpha))-\nu(0) \Big)=\frac{x_0-(\bar{x}_T-\alpha) e^{-\int^T_0r(s)ds}}{B-\bar{x}_T+\alpha}. \label{cvar_underdelta}
\end{align}
Note that the $\bar{d}$ can be computed by (\ref{def_bard}), as it is independent of $\alpha$.

Under Assumption \ref{assumption_deterministic}, let
\begin{equation}
\alpha^*:=\arg \min J(\alpha), \label{optimal_alpha}
\end{equation}
where
\begin{align}
J(\alpha):=\begin{dcases}
\alpha-\frac{\gamma}{1-\beta}( 1-\Phi(\bar{K}_2(0,\alpha))) & \hbox{if}~
\underline{d}(\alpha)<d <\bar{d},\\
\alpha-\frac{\gamma}{1-\beta}( 1-\Phi(\bar{K}_2(0,\alpha))) & \hbox{if}~d\leq \underline{d}(\alpha) ~\hbox{and}~x_0<e^{-\int_0^T r(s)ds}(\bar{x}_T-\alpha),\\
\alpha &\hbox{if}~d\leq \underline{d}(\alpha) ~\hbox{and}~x_0\geq e^{-\int_0^T r(s)ds}(\bar{x}_T-\alpha).\\
+\infty, &\hbox{otherwise}.
\end{dcases}\label{def_J}
\end{align}

\begin{cor}\label{cor_cvar}
Under Assumption \ref{assumption_deterministic}, the optimal solution of problem ($\P_{cvar}$) takes one of the following forms:
(i) If $\underline{d}(\alpha^*)<d <\bar{d}$, then $x^*(t)$ and $\pi^*(t)$ are given as
\begin{align}
\label{cor_cvar_x} x^*(t)&=e^{m(t)+\frac{\nu^2(t)}{2}}\bigg((B-\bar{x}_T+\alpha^*)\Phi\big( \bar{K}_1(t,\alpha^*)-\nu(t)\big)\\
&~~~~~~~~~~~~~~~~~~+(\bar{x}_T-\alpha^*)\Phi\big(\bar{K}_2(t,\alpha^*)-\nu(t)\big) \bigg),\notag\\
\label{cor_cvar_pi}\pi^*(t)&=\frac{e^{m(t)+\frac{\nu^2(t)}{2}}}{\sqrt{2\pi}\nu(t)} \bigg((B-\bar{x}_T+\alpha^*)e^{-\frac{(\bar{K}_1(t,\alpha^*)-\nu(t))^2}{2}}  \\
&~~~~~~~~~~~~~~~~~~+(\bar{x}_T-\alpha^*)e^{-\frac{(\bar{K}_2(t,\alpha^*)-\nu(t))^2}{2}}\bigg)
(\sigma(t)\sigma(t)^{\prime})^{-1}b(t), \notag
\end{align}
where $\bar{K}_1(t,\alpha)$ and $\bar{K}_2(t,\alpha)$ are given as
\begin{align}
\bar{K}_1(t,\alpha)=\frac{\ln\left(\frac{\lambda(\alpha)}{\eta(\alpha) z(t)}\right)-m(t)}{\nu(t)},
~\bar{K}_2(t,\alpha)= \frac{\ln\left(\frac{\lambda(\alpha)+1}{\eta(\alpha) z(t)}\right)-m(t)}{\nu(t)}, \label{cor_cvar_K}
\end{align}
with  $\lambda(\alpha)$, $\eta(\alpha)$ being the solution to the following two equations,
\begin{align}
&\label{cor_cvar_eq1}(B-\bar{x}_T+\alpha)\Phi\big(\bar{K}_1(0,\alpha)\big)
+(\bar{x}_T-\alpha)\Phi\big(\bar{K}_2(0,\alpha)\big)=d,\\
&\label{cor_cvar_eq2}(B-\bar{x}_T+\alpha)\Phi\big(\bar{K}_1(0,\alpha)-\nu(0)\big)
+(\bar{x}_T-\alpha)\Phi\big(\bar{K}_2(0,\alpha)-\nu(0)\big)\\
&~~~~~~~~~~~~~~~~~~~~~~~~~~~~~~~~~~~~=e^{\int_0^Tr(s)ds}x_0.\notag
\end{align}

(ii) If $d\leq \underline{d}(\alpha^*)$ and $x_0<e^{-\int_0^Tr(s)ds}(\bar{x}_T-\alpha^*)$, then $x^*(t)$  and $\pi^*(t)$ are given as in (\ref{cor_cvar_x}) and (\ref{cor_cvar_pi}), respectively, with $\lambda(\alpha^*)=0$ and $\eta(\alpha^*)=1/\hat{\rho}(\alpha^*)$, where $\hat{\rho}(\alpha^*)$ is given in (\ref{cvar_hatrho}).

(iii)  If $d \leq \underline{d}(\alpha^*)$ and $x_0\geq e^{-\int_0^Tr(s)ds}(\bar{x}_T-\alpha^*)$, then there are multiple optimal solutions. One of the solution is given as
\begin{align*}
x^*(t)&=e^{m(t)+\frac{\nu^2(t)}{2}}\Big((B-\bar{x}_T+\alpha^*)\Phi\big(K_3(t,\alpha^*)
-\nu(t)\big)+\gamma \Big)\\
\pi^*(t)&=\frac{1}{\sqrt{2\pi}\nu(t)}e^{m(t)+\frac{\nu^2(t)}{2}}
(B-\bar{x}_T+\alpha^*)e^{-\frac{(K_3(t,\alpha^*)-2\nu(t))^2}{2}}
\big(\sigma(t)\sigma(t)^{\prime}\big)^{-1}b(t),
\end{align*}
where ${K}_3(t):=\big(\ln\big(\underline{\delta}(\alpha^*)/z(t)\big)-m(t)\big)/\nu(t)$ with $\underline{\delta}(\alpha^*)$ being the solution of (\ref{cvar_underdelta}).
\end{cor}

\proof For any fixed $\alpha$, problem $(\P_{cvar}(\alpha))$ takes the same form as $(\P_{lpm}^1)$. Thus, for case (i), substituting $\gamma$ by $\bar{x}_T-\alpha$ to (\ref{thm_lpm_dq<1_x}), (\ref{thm_lpm_dq<1_pi}), (\ref{thm_lpm_dq<1_eq1}) and (\ref{thm_lpm_dq<1_eq2}) gives rise to the results in (\ref{cor_cvar_x}), (\ref{cor_cvar_pi}), (\ref{cor_cvar_eq1}) and (\ref{cor_cvar_eq2}), respectively. For case (ii) and (iii), we just substitute $\gamma$ by $\bar{x}_T-\alpha$ in (ii) and (iii) of Theorem \ref{thm_lpm_dq<1}, respectively. The object value in (\ref{thm_lpm_dq<1_obj}) then becomes
\begin{align*}
E[(\bar{x}_T-\alpha-x^*(T))_+]=(\bar{x}_T-\alpha)(1-\Phi(\bar{K}_2(0,\alpha))).
\end{align*}
From Lemma \ref{lem_cvar}, we know that $\CVaR[f(x(T))]$ can be computed by minimizing $\alpha$ in $(\ref{def_par_cvar})$. It can be verified that $J(\alpha)$ defined in (\ref{def_J}) is $\alpha+\frac{1}{1-\beta}\E[(\bar{x}_T-x^*(T)-\alpha)_+]$.

\endproof

Note that the optimal $\alpha^*$ defined in (\ref{optimal_alpha}) may not be unique. Due to the special feature of the distribution function of $x^*(T)$ , from Theorem 10 in \cite{Rockafellar:2002}, the set $\{\alpha~|~\alpha=\arg \min_{\alpha}J(\alpha)\}$  is a closed and bounded interval.
Since the objective function $J(\alpha)$ is convex with respect to $\alpha$, we can use the following gradient searching procedure to find one optimal $\alpha^*$ in Corollary \ref{cor_cvar}.

\noindent \underline{{\bf Searching algorithm for $\alpha^*$ }}

\indent \underline{Input}: The parameters of problem $(\P_{cvar})$, small positive numbers $\epsilon>0$ and $\zeta>0$, the step size $\vartheta>0$.

\indent \underline{Step 0} Choose $\alpha\leftarrow\alpha_0$ as the initial point and small positive number $\epsilon>0$ as the stopping criteria. Go to Step 1.

\indent \underline{Step 1} For given $\alpha$, let $\hat{\alpha}\leftarrow \alpha+\zeta$, then compute $J(\alpha)$ and $J(\hat{\alpha})$ by (\ref{def_J}). Go to the next step.

\indent \underline{Step 2} Compute the gradient
$\kappa=\big( J(\hat{\alpha})-J(\alpha) \big)/\zeta$. If $|\kappa|<\epsilon$, return $\alpha$ as the optimal solution. Otherwise, let $\alpha=\alpha+ \vartheta\cdot \kappa$. Go to Setp 1.

Note that when implementing the above gradient searching procedure, controlling the step size plays a key role. Furthermore, the above procedure can only guarantee identification of one optimal solution of $\alpha^*$.

%
%\noindent \underline{{\bf Searching algorithm for $\alpha^*$ in set $\mathcal{S}$}}
%
%\indent \underline{Input}: The parameters of problem $(\P_{cvar})$, small positive numbers $\epsilon>0$ and $ $, the step size $\varrho>0$.
%
%\indent \underline{Step 0} Choose $\alpha\leftarrow\alpha_0$ as the initial point and small positive number  $\epsilon>0$ as the stopping criteria. Go to Step 1.
%
%\indent \underline{Step 1} For given $\alpha$, let $\hat{\alpha}\leftarrow \alpha+\delta$ and compute $\underline{d}(\alpha)$ and $\underline{d}(\hat{\alpha})$.
%
%\indent \underline{Step 2} If $\underline{d}(\alpha)<d$, solve (\ref{cor_cvar_eq1}) and (\ref{cor_cvar_eq2}) for $\lambda(\alpha)$, $\eta(\alpha)$, otherwise, let $\lambda(\alpha)=0$ and solve $\eta(\alpha)$ from (\ref{cor_cvar_eq2}). Repeat this step for $\hat{\alpha}$ to get $\lambda(\hat{\alpha})$, $\eta(\hat{\alpha})$.
%
%\indent \underline{Step 3} Compute the objective value $g(\alpha)$ and $g(\hat{\alpha})$ from (\ref{cor_cvar_obj}) with the gradient $\kappa=\big(g(\hat{\alpha})-g(\alpha) \big)/\delta$.  If $|\kappa|<\epsilon$, return $\alpha$ as the optimal solution. Otherwise, let $\alpha=\alpha+ \xi\cdot \kappa$. Go to Setp 1.
%
%Note that the above procedure can only guarantee identification of one optimal solution of $\alpha^*$.

%*******************************************************************************
\section{Illustrative Examples and Comparison}\label{se_example}
In this section, we first investigate an illustrative example to compare the dynamic mean-downside risk portfolio policy derived in this paper with the well known dynamic mean-variance portfolio policy. The continuous-time dynamic mean-variance portfolio selection problem is solved by \cite{ZhouLi:2000} for the market setting with a deterministic opportunity set, by \cite{Lim:2002} for the case with a stochastic opportunity set, and by \cite{Bielecki:2005} for the case with bankruptcy restriction. In this section we compare our results with the one in \cite{Bielecki:2005}, in which no bankruptcy restriction is placed as we do in this paper for our mean-downside risk portfolio models.

Let us discuss first the solution to the following dynamic mean-variance portfolio optimization model,
\begin{align*}
(\P_{mv})~\min_{\pi(\cdot)\in \L^2_{\F}(0,T; \R^n) }~&~\textrm{var}[x(T)]:= \E[x(T)^2]-(\E[x(T)])^2\\
\textrm{Subject to}~&~
                      \begin{dcases}
                        \E[x(T)]=d, \\
                        \textrm{($x(\cdot)$,$\pi(\cdot)$) statisfies dynamics (\ref{def_wealth}) }, \\
                        0 \leq x(T).
                        %\pi(\cdot) \in \L_{\F}^2(0,T;\R^n),
                      \end{dcases}%\label{def_P1_constraint}
\end{align*}
Different from \cite{Bielecki:2005} in which Bielecki et al. introduce a fictitious security to represent the optimal wealth process and optimal portfolio policy, we represent the optimal wealth process and portfolio policy in terms of the state price density $z(t)$. Actually, $z(t)$ can be also regarded as an artificial security. Although these two ways are equivalent, we modify the result in \cite{Bielecki:2005} to fit our purpose of comparison.

We still use the martingale approach to solve problem ($\P_{mv}$) and find the optimal terminal wealth by solving the following auxiliary problem,
\begin{align*}
(\A_{mv})~&~\min_{X\in \L_{\F_T}^2(\Omega,\R) }~\E[X^2]-d^2, \\
\textrm{Subject to}~&~
                      \begin{dcases}
\E[X] = d, \\
\E[z(T)X]=x_0,  \\
0 \leq X.
\end{dcases}
\end{align*}

\begin{thm}\label{thm_mv}
(i) The optimal terminal wealth of problem $(\A_{mv})$ is
\begin{align}
X^*=\frac{1}{2}(\lambda-\eta z(T))\1_{\lambda-\eta z(T)\geq 0}, \label{def_mv_X}
\end{align}
where the parameters $\lambda>0$ and $\eta>0$ are the solution to the following system of two equations,
\begin{align}
&\E[(\lambda-\eta z(T))_+]=2d,\\
&\E[x(T)(\lambda-\eta z(T))_+]=2x_0.
\end{align}

(ii) Under Assumption \ref{assumption_deterministic}, the optimal wealth process and optimal portfolio policy of $(\P_{mv})$ are given, respectively, as
\begin{align}
\label{mv_x} x^*(t)&=\frac{\lambda}{2} e^{m(t)+\frac{\nu^2(t)}{2}}\Phi(K(t)-\nu(t))-\frac{\eta}{2} z(t) e^{2m(t)+2\nu^2(t)}\Phi(K(t)-2\nu(t)),\\
\label{mv_pi}\pi^*(t)&=\Big( \frac{\lambda }{2\nu(t)\sqrt{2\pi}}e^{m(t)+\frac{\nu^2(t)}{2}-\frac{(K-\nu(t))^2}{2}}
      -\frac{\eta z(t)}{2}e^{2m(t)+2\nu^2(t)}\Big(\Phi\big(K(t)-2\nu(t)\big)\\
      &~~~~~~-\frac{1}{\sqrt{2\pi}\nu(t)} e^{-\frac{(K(t)-2\nu(t))^2}{2}} \Big)\Big)(\sigma(t)\sigma(t)^{\prime})^{-1}b(t),\notag
      \end{align}
where $K(t)=\ln(\lambda/\eta-m(t))/\nu(t)$, and $m(t)$ and $\nu(t)$ are  defined in (\ref{def_m(t)}) and (\ref{def_v(t)}), respectively. Furthermore, parameters $\eta$ and $\lambda$ are the solution to the following two equations,
\begin{align}
&\lambda \Phi(K(0))-\eta e^{m(0)+\frac{\nu^2(0)}{2}}\Phi(K(0)-\nu(0))=2d,\label{mv_eq1}\\
&\lambda e^{m(0)+\frac{\nu^2(0))}{2}}\Phi(K(0)-\nu(0))-\eta e^{2m(0)+2\nu^2(0)}\Phi(K(0)-2\nu(0))=2x_0.\label{mv_eq2}
\end{align}

\end{thm}

\proof The proof of result (i) can be found in \cite{Bielecki:2005} and the result (ii) can be proved by a method similar to the proof in Theorem \ref{thm_lpm_dq>1}.
\endproof

\begin{exam}\label{exam_compare}
We consider the following example to demonstrate the properties of the mean-LPM problem, $(\P_{lpm}^{q})$,  with all the market parameters being set as the same as in Example 7.1 of \cite{Jin:2005}. The risk free rate is $r(t)=0.06$ and there is only one risky asset with $\mu(t)=0.12$ and $\sigma(t)=0.15$ for $t\in [0,T]$. The initial wealth is $x(0)=1$ (in a unit of thousand dollars, for example), the expected terminal payoff is $d=1.3$, and the investment horizon is $T=1$ year. The benchmark level is set as $\gamma=e^{0.06}x(0)=1.0618$, which is the payoff of the investment solely in the bank account. We also set the upper bound of terminal wealth as  $B=10$. Now, we compare the mean-LPM portfolio optimization models, $(\P_{lpm}^2)$ and $(\P_{lpm}^1)$ with the mean-variance portfolio model $(\P_{mv})$.
We first compute the parameters $\underline{d}$ and $\bar{d}$ by Proposition \ref{prop_d} (see Table \ref{table_parameter}). By using Theorem \ref{thm_lpm_dq>1} and Theorem \ref{thm_lpm_dq<1}, we solve the pair of Lagrange multipliers, $\eta$ and $\lambda$,  for problem $(\P_{lpm}^2)$ according to (\ref{thm_lpm_dq>1_eq1}) and (\ref{thm_lpm_dq>1_eq2}); for problem $(\P_{lpm}^1)$  according to (\ref{thm_lpm_dq<1_eq1}) and (\ref{thm_lpm_dq<1_eq2}); and for problem $(\P_{mv})$ according to (\ref{mv_eq1}) and (\ref{mv_eq2}), respectively, which are listed in Table \ref{table_parameter}. Following Theorems \ref{thm_lpm_dq>1}, \ref{thm_lpm_dq<1} and \ref{thm_mv}, we can also compute the analytical expressions of optimal wealth $x^*(t)$ and optimal portfolio policy $\pi^*(t)$ for problems $(\P_{lpm}^{2})$, $(\P_{lpm}^{1})$ and $(\P_{mv})$, respectively.

\begin{table}
  \centering
  \begin{tabular}{ccccc}
    \toprule
                 & $\eta$  & $\lambda$   & $\underline{d}$  & $\bar{d}$\\
\hline
${\P}_{lpm}^{2}$ & $0.2007$ & $0.7852$ & $1.0618$ & $1.9847$\\
${\P}_{lpm}^{1}$ & $0.7852$ & $0.3261$ & $1.0618$ & $1.9847$\\
$\P_{mv}$            & $3.421 $ & $5.7694$ & -        & -      \\
\bottomrule
\end{tabular}
\caption{Parameters $\lambda$ and $\eta$ in Example \ref{exam_compare}}\label{table_parameter}
\end{table}

Figures \ref{fig:exam1_xT} and \ref{fig:exam1_xt} show the optimal wealth $x^*(t)$ of problems ($\P^2_{lpm}$), ($\P^{1}_{lpm}$) and $(\P_{mv})$ at $t=1$ and $t=0.5$, respectively. We can see that when $z(t)$ is small, i.e., the market condition is good, the wealth level $x^*(t)$ of problem $(\P_{lpm}^{2})$ (or $(\P_{lpm}^1)$) is much higher than the wealth level generated from problem ($\P_{mv}$). As $z(t)$ increases, i.e., the market condition becomes worse, the wealth level of $(\P_{mv})$ reduces to zero faster than the wealth level of $(\P_{lpm}^{2})$ or ($\P_{lpm}^1$). As for the portfolio policy $\pi^*(t)$, which is plotted in Figure \ref{fig:exam1_pi}, we can see that the mean-variance policy of ($\P_{mv}$) allocates most wealth for the intermediate range of state $z(t)$. However, the mean-LPM policies of $(\P_{lpm}^{2})$ and $(\P_{lpm}^1)$ allocate more wealth in the risky asset when the market state is in a good condition. When the market condition is in the mediant state, a mean-LPM investor tends to allocation his wealth in risk free asset. Contrary to intuitive thinking, when the market condition becomes worse (i.e., $z(t)$ is increasing), the mean-LMP portfolio policy increases its demand in the risky asset. Figure \ref{fig:exam1_xwt} plots the relationship between the proportion in the risky asset, $w^*(t)=\pi^*(t)/x^*(t)$, and $x^*(t)$, and demonstrates a feature of the threshold type, i.e., there is a threshold around $1$ under which or above which the LPM investor increases his allocation in the risky asset.  Compared with the ($\P_{lpm}^2$) policy, the ($\P_{lpm}^1$) policy is more  aggressive when the current wealth $x(t)$ is below the threshold, and a similar pattern appears between them when the wealth is above the threshold. This kind of feature is significantly different from the mean-variance policy and the policy generated from the utility maximization. Figures \ref{fig:exam1_lpm1_xw} and \ref{fig:exam1_lpm2_xw} show the allocations in the risky asset for different time points $t=0.2$, $t=0.5$ and $t=0.8$. Generally speaking, as the investment approaches to the terminal time, the mean-LPM policies increase their allocations in the risk assets. However, when the wealth is around the threshold point, the $(\P_{lpm}^1)$ policy is more sensitive to the time than the $(\P_{lpm}^2)$ policy. We can also compute the probability that $x^*(T)$ reaches the upper bound $B=10$ as $2.2\%$ and $3.2\%$, respectively, for $(\P_{lpm}^1)$ and $(\P_{lpm}^2)$. If we increase $B$ to $30$, then the probability drops to $0.7\%$ and $0.9\%$, respectively, for $(\P_{lpm}^1)$ and $(\P_{lpm}^2)$. That is to say, although there is an upper bound on the wealth level in problems ($\P_{lpm}^1$) and ($ \P_{lpm}^2$), the probability that the wealth level actually reaches such an upper bound is very small. Figure \ref{fig:exam1_B} plots the $(x^*(t), w^*(t))$ pair for different values of $B$. We can see that when the current wealth level $x^*(t)$ is above the threshold, allocation to the risky asset becomes more aggressive when the upper bound $B$ is increasing. However, when the current wealth is below or near the threshold, both optimal policies of $(\P_{lpm}^1)$ and $(\P_{lpm}^2)$  keep almost invariant with respect to $B$. We can thus conclude that, although the upper limit $B$ affects the investment policy, the portfolio weight is actually quite robust with respect to $B$, if the current wealth does not deviate too much from the threshold.

%\begin{align*}
%x^*(T)&=(8.9382+0.0533z(T)-0.0191)\1_{z(T)\leq 0.3582}\\
%&~~~+(1.0618-0.0533z(T)+0.0191)\1_{z(T)\leq 20.298},\\
%x^*(t)&=
%\end{align*}
%(ii) If we consider problem $(\bar{P}_{lpm})$ with $p=1$, similarly, we can solve $\eta=0.6360$ and  $\lambda=0.2391$ from the equations (\ref{thm_lpm_dq<1_eq1}) and (\ref{thm_lpm_dq<1_eq2}). The optimal terminal wealth is
%\begin{align*}
%x^*(T)=8.9382\1_{z(T)\leq 0.3759}+1.0618\1_{z(T)\leq 1.9483}.
%\end{align*}
%(iii) As a benchmark, we can also solve the mean-variance portfolio optimization model ($\P_{mv}$) by using Theorem \ref{thm_mv}, which yields $\eta =1.75$ and $\lambda=4.0441$ with the optimal terminal wealth being
%\begin{align*}
%x^*(T)=(2.022-0.875z(T))\1_{z\leq 2.31}.
%\end{align*}
%The Figure \ref{fig_lpm_mv1} and \ref{fig_lpm_mv2} shows the optimal terminal wealth of the problem $(\bar{\P}_{lpm})$ with $p=2$ and $p=1$ comparing with the optimal terminal wealth of the  mean-variance portfolio optimization model. We can see from Figure \ref{fig_lpm_mv1} and \ref{fig_lpm_mv2} that when $z(T)$ is small the optimal termial

%************************ lpm p=2 ***********************************************
\begin{figure}[hbpt]
\centering
\subfigure[Optimal wealth $x^*(T)$ at $T=1$]{
\label{fig:exam1_xT} %% label for first subfigure
\includegraphics[width=0.46\textwidth]{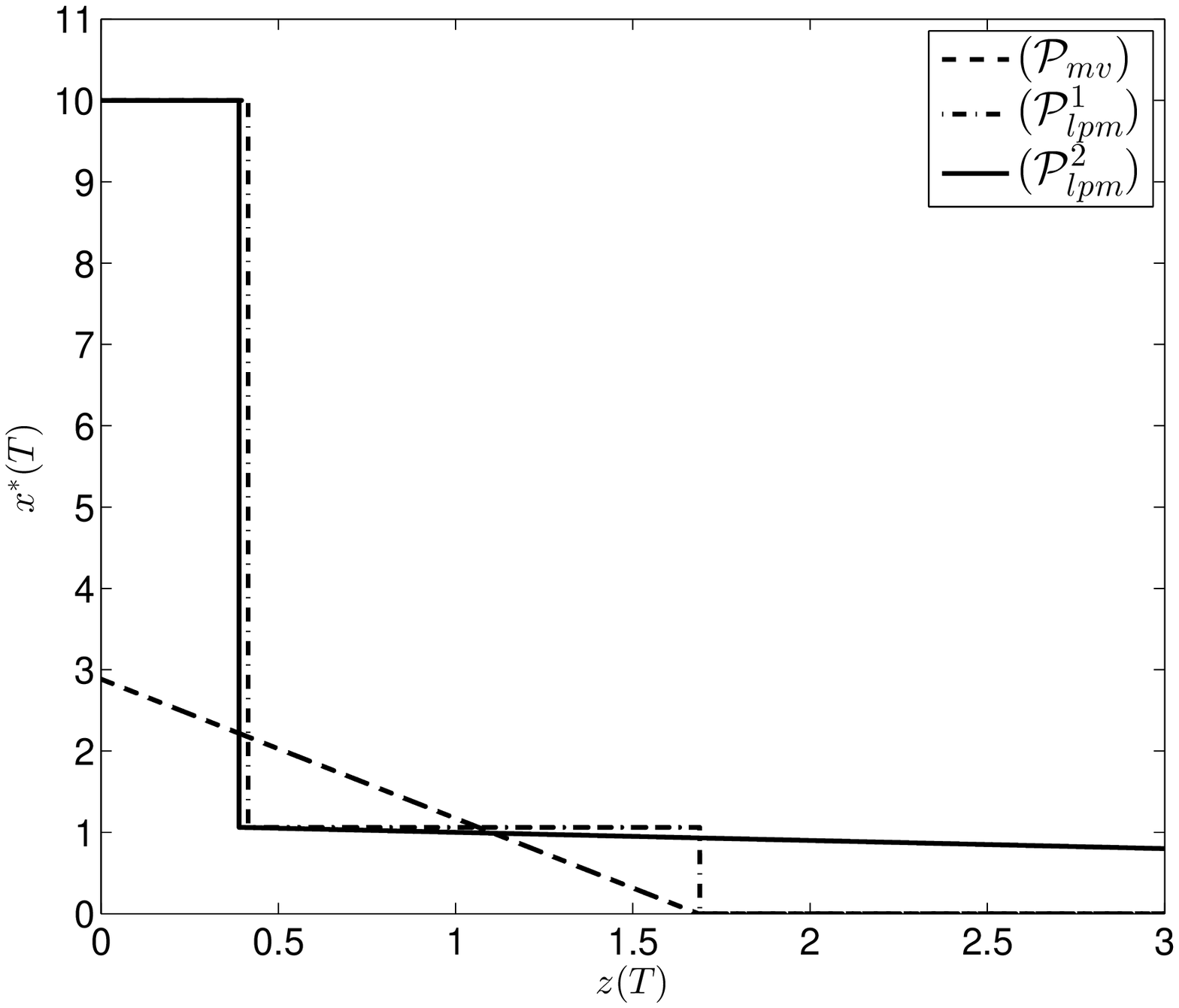}}
\subfigure[Optimal wealth $x^*(t)$ at $t=0.5$]{
\label{fig:exam1_xt}
\includegraphics[width=0.46\textwidth]{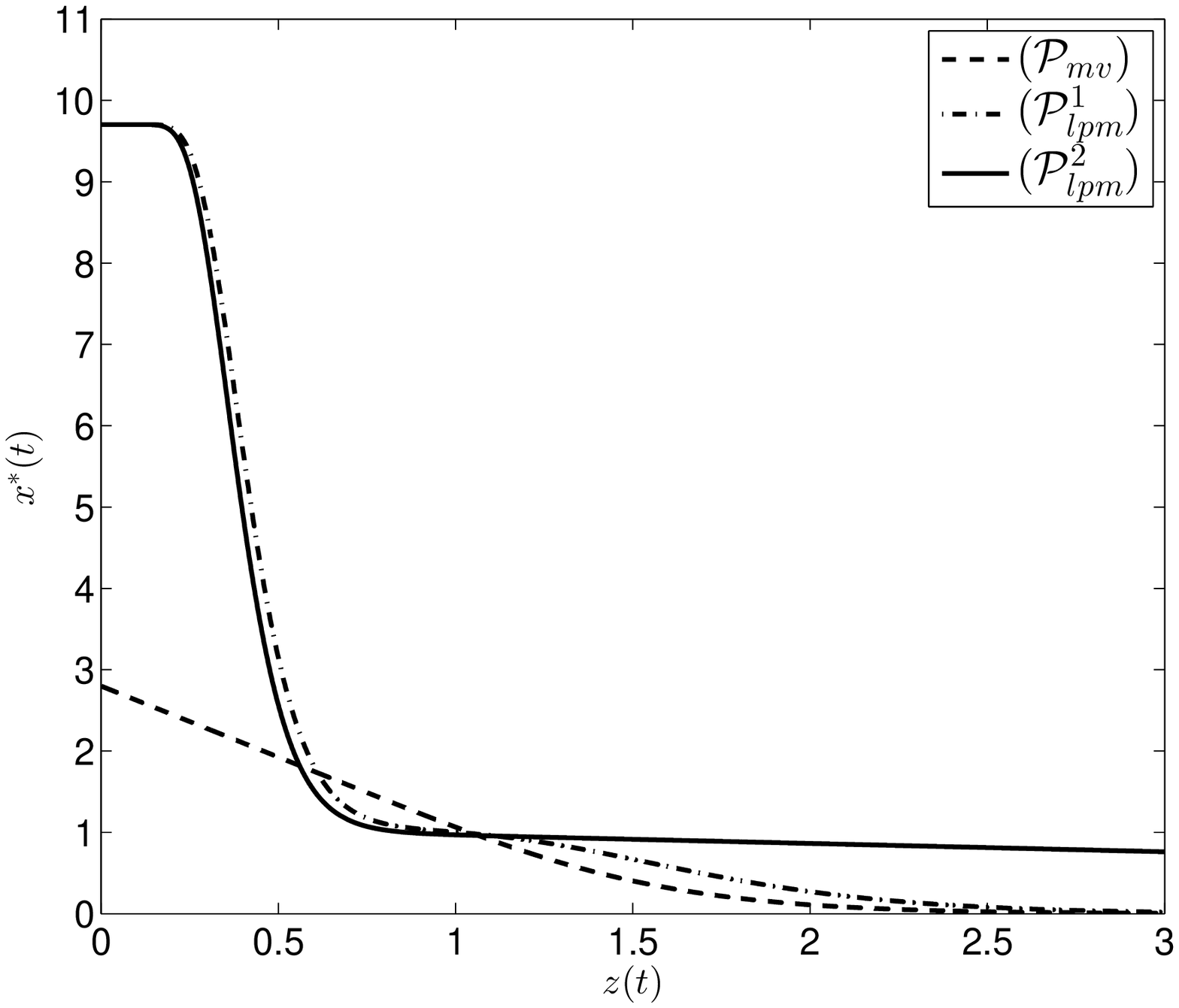}}
\subfigure[Optimal portfolio $\pi^*(t)$ at $t=0.5$]{
\label{fig:exam1_pi}
\includegraphics[width=0.46\textwidth]{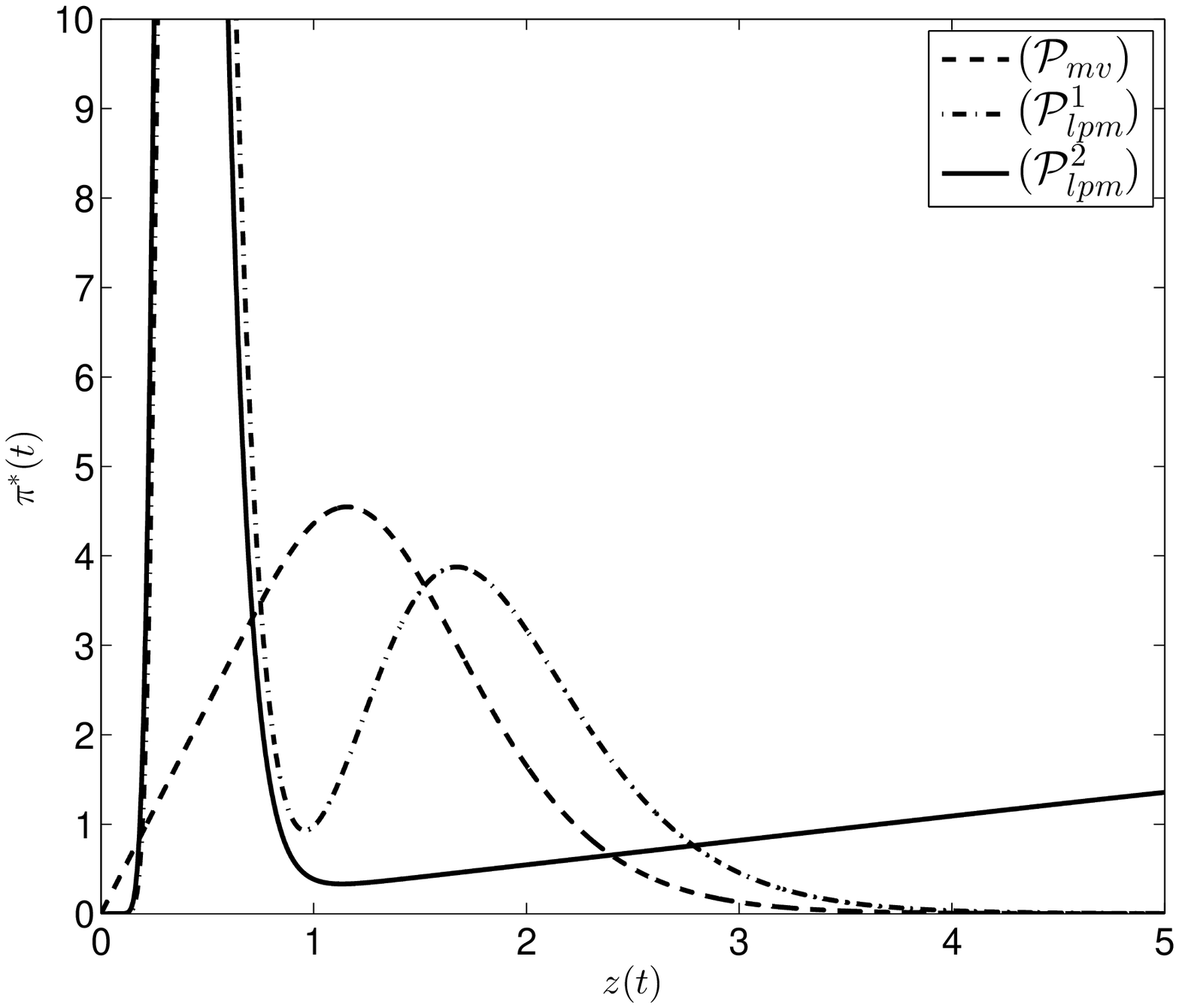}}
\subfigure[The optimal portfolio and wealth pair $(x^*(t),\pi^*(t))$ at $t=0.5$]{
\label{fig:exam1_xwt}
\includegraphics[width=0.46\textwidth]{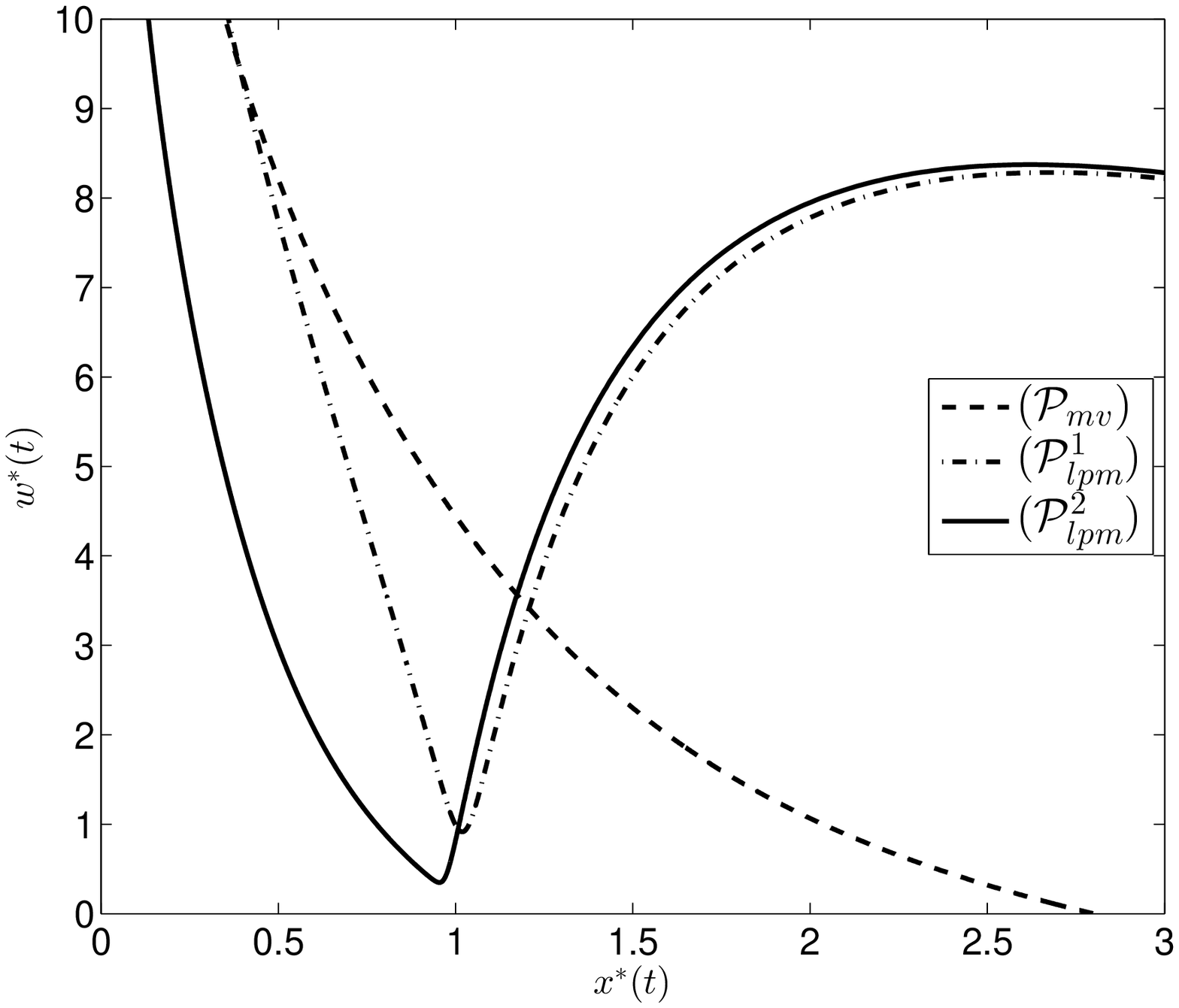}}
\caption{Optimal wealth and portfolio of problems ($\P_{mv}$), ($\P_{lpm}^1$) and ($\P_{lpm}^2$) for Example \ref{exam_compare}} \label{fig:exam1_xuw}
\end{figure}

\begin{figure}[hbpt]
\centering
\subfigure[The $(w^*(t),x^*(t))$ pair of $(\P_{lpm}^1)$]{
\label{fig:exam1_lpm1_xw} %% label for first subfigure
\includegraphics[width=0.46\textwidth]{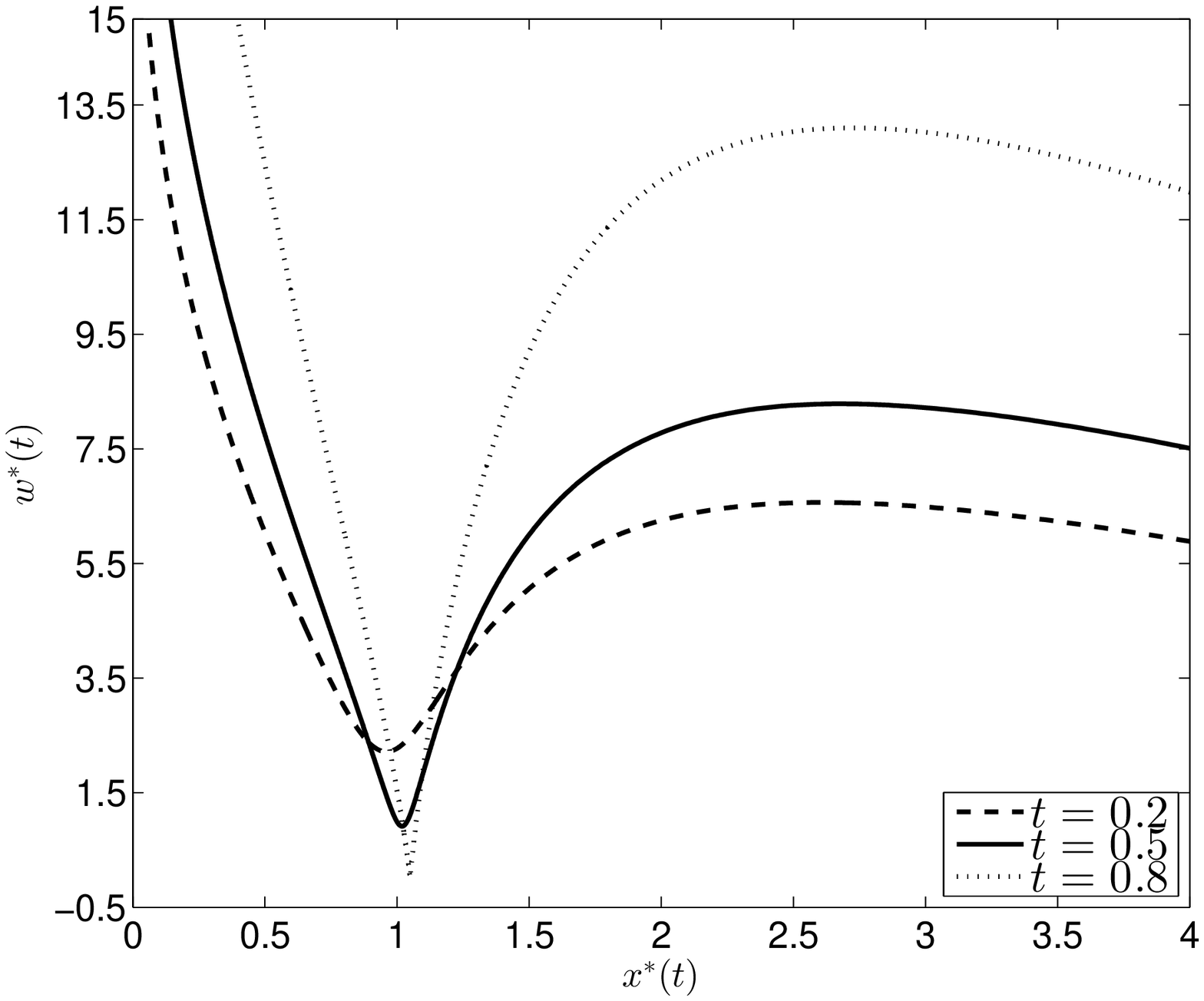}}
\subfigure[The $(w^*(t),x^*(t))$ pair of $(\P_{lpm}^2)$]{
\label{fig:exam1_lpm2_xw}
\includegraphics[width=0.46\textwidth]{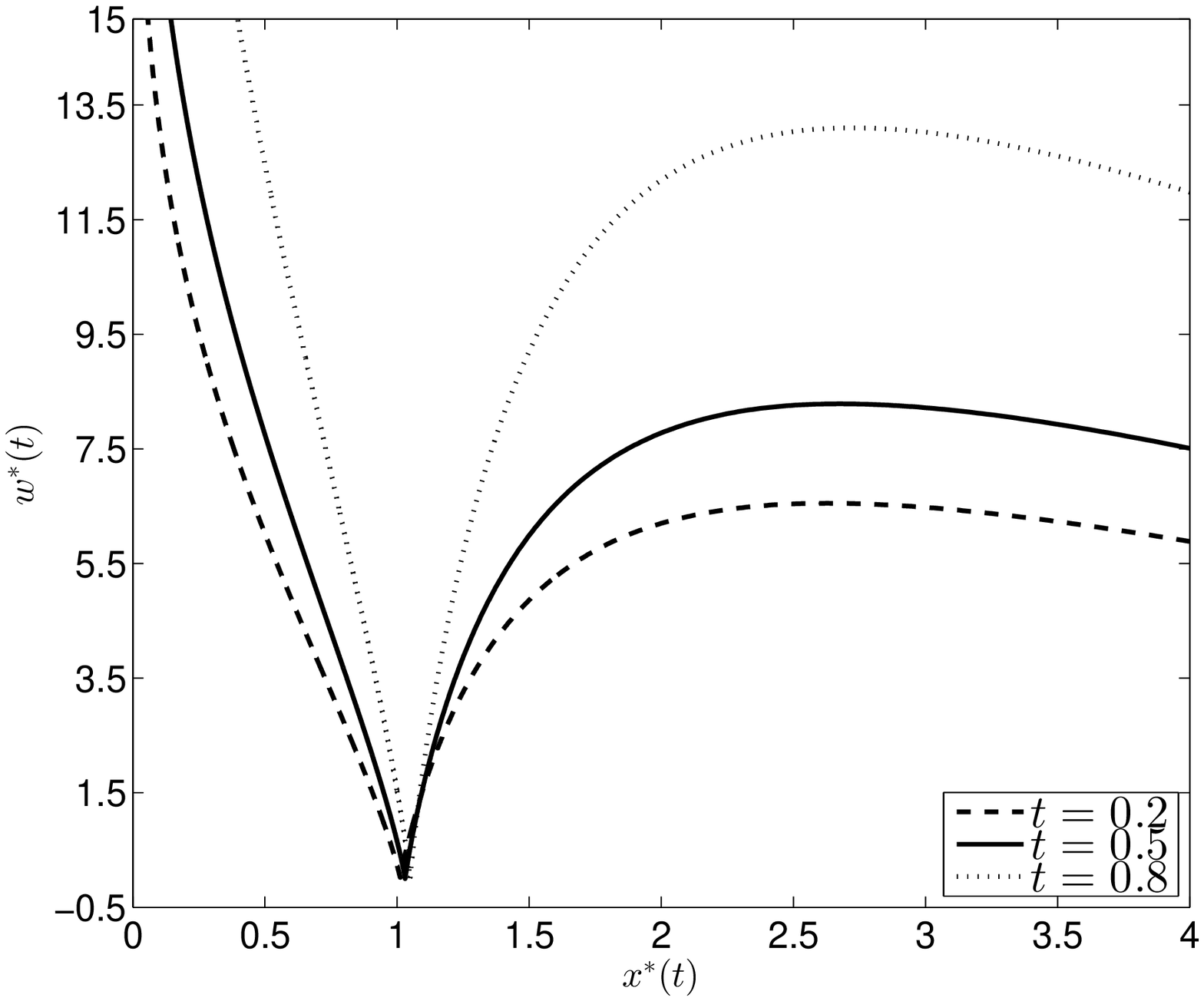}}
\caption{The optimal  portfolio policy pair $(w^*(t),x^*(t))$ of problems ($\P_{lpm}^1$) and ($\P_{lpm}^2$) in  Example \ref{exam_compare} at $t=0.2$, $t=0.5$ and $t=0.8$}\label{fig:exam1_time}
\end{figure}

\begin{figure}[hbpt]
\centering
\subfigure[The $(w^*(t),x^*(t))$ pair of $(\P_{lpm}^1)$]{
\label{fig:exam1_lpm1_diffB} %% label for first subfigure
\includegraphics[width=0.46\textwidth]{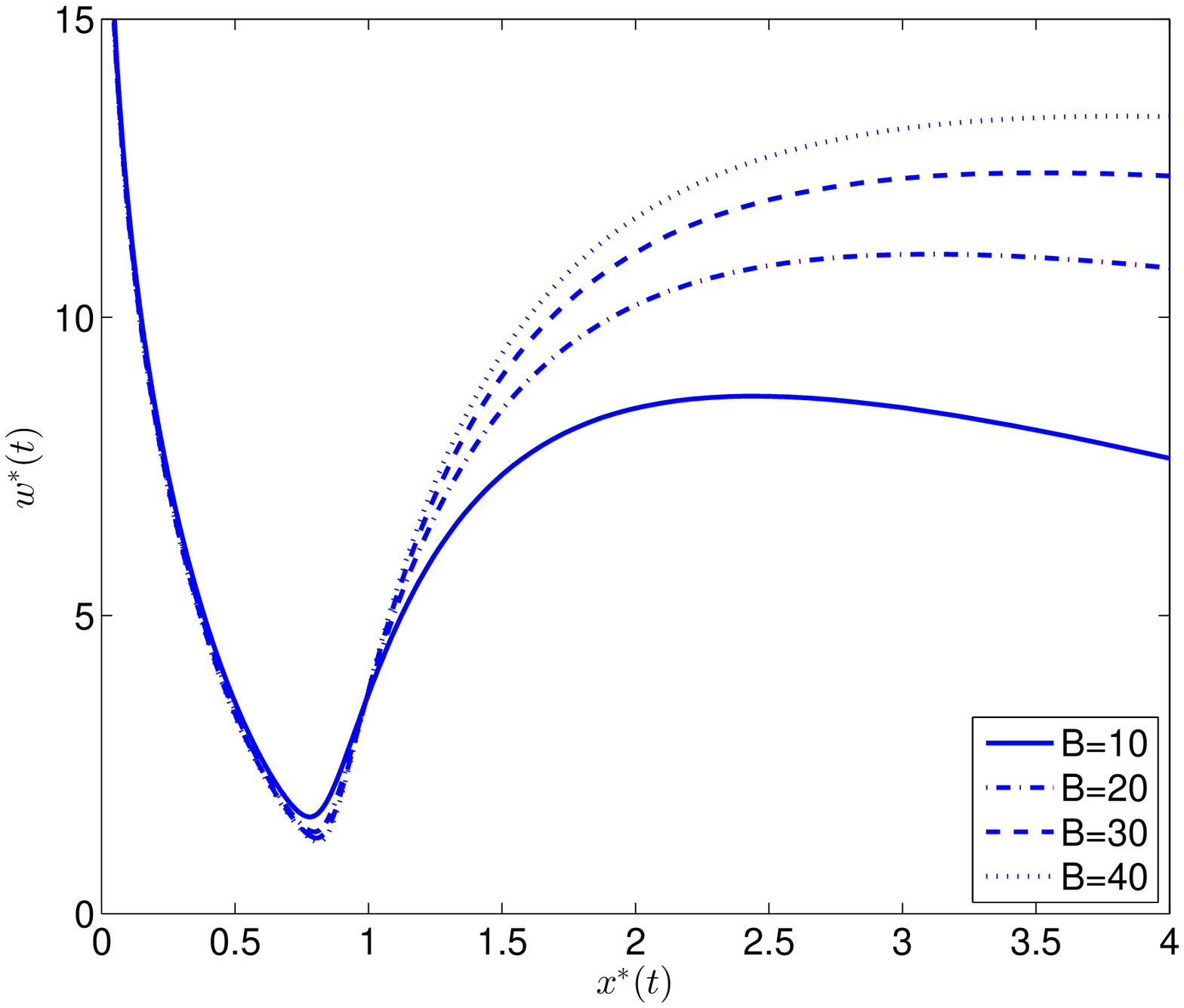}}
\subfigure[The $(w^*(t),x^*(t))$ pair of $(\P_{lpm}^2)$]{
\label{fig:exam1_lpm2_difB}
\includegraphics[width=0.46\textwidth]{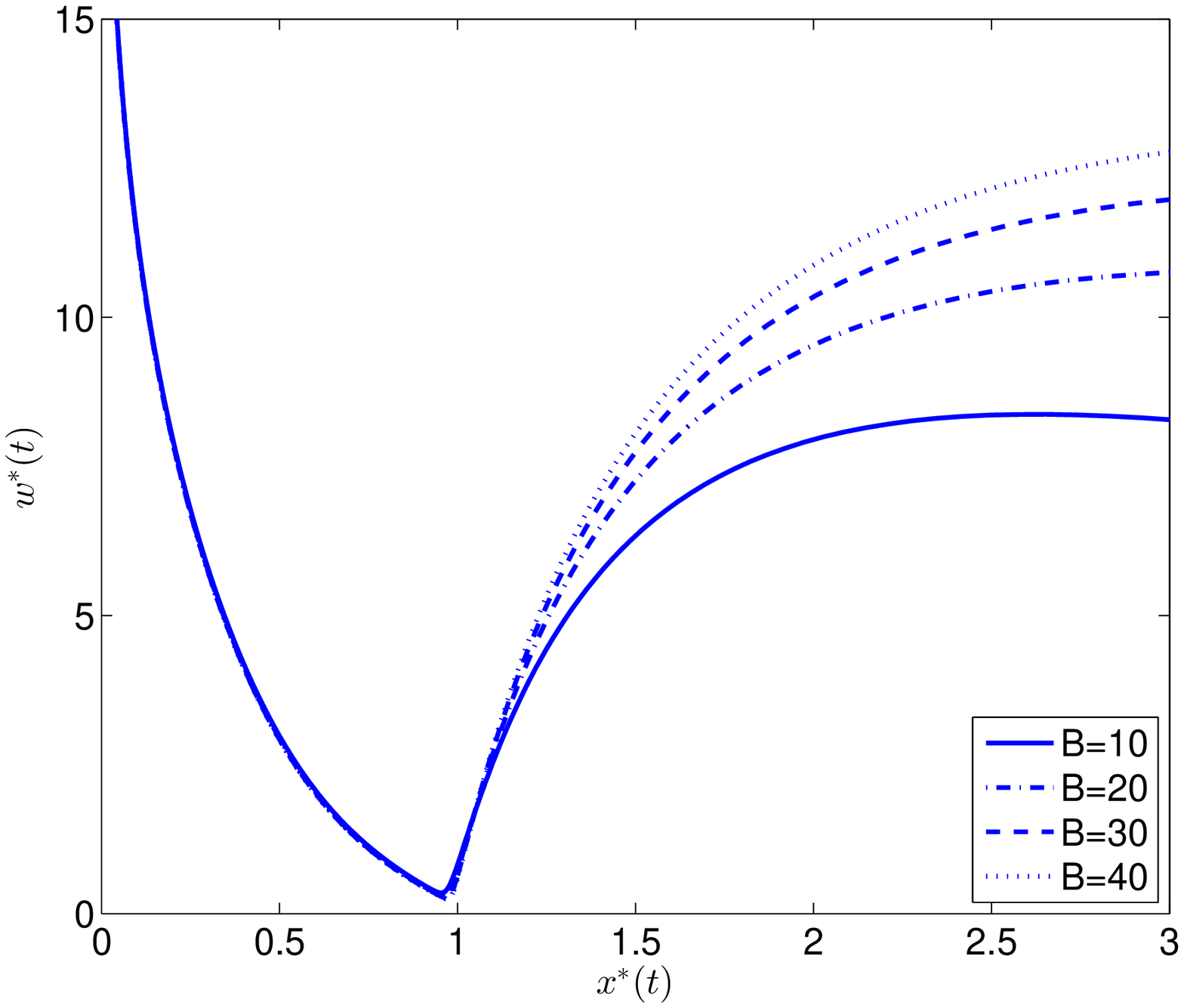}}
\caption{The optimal  portfolio policy pair $(w^*(t),x^*(t))$ of problems ($\P_{lpm}^1$) and ($\P_{lpm}^2$) in  Example \ref{exam_compare} at $t=0.5$ for different $B$}\label{fig:exam1_B}
\end{figure}

\end{exam}

%************************* example CVaR ************************
\begin{exam}\label{exam_cvar}
In this example, we compare the dynamic mean-CVaR portfolio optimization model studied in Section \ref{se_cvar} with the well known static mean-CVaR portfolio model proposed in \cite{Rockafellar:2000} and \cite{Rockafellar:2002}. We adopt a market setting similar to that given in \cite{Rockafellar:2000}, where the portfolio is constructed by three assets, the Standard $\&$ Poor 500 index (S$\&$P 500), the long-term US Government Bond (Bond), and the portfolio of the US small capital stocks (Small-Cap). We scale the statistics (Tables 1 and 2 in \cite{Rockafellar:2000}) of the monthly returns listed in \cite{Rockafellar:2000} to the annual ones. Table \ref{table_exam_cvar} lists the mean value and the covariance of the asset returns. Note that the expected return rate and the covariance matrix are estimated by using the sample mean and sample covariance. Different from the assumption in \cite{Rockafellar:2000}, we assume that the asset returns are log-normally distributed instead of normally distributed, as we assume in this study that the assets prices follow the SDE in (\ref{def_risky_sde}), from which the resulted distributions of the asset returns are indeed log-normally distributed when the market parameters are deterministic. We assume that the drift rate vector $\mu(t)$ and volatility matrix $\sigma(t)$ are constants, i.e., $\mu(t)=\mu$ and $\sigma(t)=\sigma$, for all $t\in[0,T]$. From Table \ref{table_exam_cvar}, we can compute parameters $\mu$ and $\sigma$ as follows,
\begin{align*}
\mu=\left(
      \begin{array}{c}
        0.1346 \\
        0.0530 \\
        0.1722 \\
      \end{array}
    \right), ~~\sigma=\left(
                        \begin{array}{ccc}
                          0.1428 & 0.0094 & 0.1002 \\
                          0.0094 & 0.0728 & 0.0031 \\
                          0.1002 & 0.0031 & 0.2353 \\
                        \end{array}
                      \right).
\end{align*}

\begin{table}
  \centering
    \begin{tabular}{ccccc}
      \hline\hline
      Assets          & Expected rate  & \multicolumn{3}{c}{ Assets Covariance } \\
      \cline{3-5}
                     & of Return  & S$\&$P 500 & Bond & Small Cap \\
      \hline
      S$\&$P 500     & $0.1213$   & 0.039  & 0.0028  & 0.0504\\
      Bond          & $0.0522$    & 0.0028 & 0.006   & 0.0023\\
      Small Cap     & $0.1645$    & 0.0504 & 0.0023  & 0.0917\\
      \hline\hline
    \end{tabular}
  \caption{The statistics of the annual returns of the assets}\label{table_exam_cvar}
\end{table}

\begin{table}
  \centering
  \begin{tabular}{ccccccc}
    \toprule
          & \multicolumn{3}{c}{Buy-and-hold Policy} &  \multicolumn{3}{c}{Dynamic Policy of $(\P_{cvar})$}  \\
    &\multicolumn{3}{c}{\CVaR(f(x(T)))} & \multicolumn{3}{c}{\CVaR(f(x(T)))}\\
  \cmidrule(r){2-4} \cmidrule(r){5-7}
     $d$ & $\beta=0.9$ & $\beta=0.95$  & $\beta=0.99$ &  $\beta=0.9$ & $\beta=0.95$ &  $\beta=0.99$ \\
\cmidrule(r){2-4} \cmidrule(r){5-7}
 $11.00$ &  $1.129$ & $1.414$ & $2.003$ & $0.056$ & $0.074$ & $0.078$ \\
$11.20$ &  $1.351$ & $1.718$ & $2.394$ & $0.079$ & $0.098$ & $0.148$ \\
$11.40$ &  $1.618$ & $2.034$ & $2.756$ & $0.104$ & $0.123$ & $0.238$ \\
$11.60$ &  $1.870$ & $2.300$ & $3.212$ & $0.130$ & $0.150$ & $0.347$ \\
$11.80$ &  $2.042$ & $2.649$ & $3.589$ & $0.158$ & $0.179$ & $0.473$ \\
$12.00$ &  $2.319$ & $2.849$ & $3.997$ & $0.187$ & $0.208$ & $0.615$ \\
$12.20$ &  $2.434$ & $3.227$ & $4.377$ & $0.218$ & $0.239$ & $0.774$ \\
$12.40$ &  $2.752$ & $3.437$ & $4.850$ & $0.249$ & $0.271$ & $0.948$ \\
$12.60$ &  $2.989$ & $3.809$ & $5.150$ & $0.282$ & $0.304$ & $1.139$ \\
$12.80$ &  $3.252$ & $3.939$ & $5.557$ & $0.316$ & $0.338$ & $1.346$ \\
$13.00$ &  $3.405$ & $4.342$ & $5.884$ & $0.351$ & $0.373$ & $1.570$ \\
    \bottomrule
  \end{tabular}

  \caption{Comparison between Buy-and-Hold policy and dynamic policy}\label{table_exam_cvar2}
\end{table}

In this example, we assume that the market is complete, which further implies that the market price of risk is $\theta(t)=\left(\begin{array}{c}
0.4864,
                                                                   0.4269,
                                                                   0.4510
                                                                    \end{array}
                                                                  \right)^{\prime}
$ for all $t\in[0,T]$. The dynamic mean-CVaR portfolio policy can be computed according to Corollary \ref{cor_cvar}. For the static buy-and-hold policy, we use the Monte Calo simulation approach (see,  e.g., \cite{Rockafellar:2000}) to compute the CVaR value, $\CVaR[f(x(T))]$. More specifically, we first randomly generate $10^5$ samples of the returns of the three assets from the log-normal distribution according to the mean and covariance listed in Table \ref{table_exam_cvar}. Note that, while the static optimization model includes the same constraints as in problem $(\P_{cvar})$, its optimal portfolio is only sought within the buy-and-hold type. We can compute the CVaR value of the buy-and-hold policy by solving the linear programming problem associated with these samples. In this example, we use CPLEX 12.3 as the solver for the correspondent linear programming problem (see, e.g., \cite{Rockafellar:2000}).

Table \ref{table_exam_cvar2} compares the CVaR values between the static buy and hold policy and our dynamic policy resulted from solving problem $(\P_{cvar})$. For different confidential levels of $\beta$ (= $0.9$, $0.95$, $0.99$) and different levels of the target terminal wealth $d$, we can observe that the dynamic mean-CVaR portfolio policy always reduces the CVaR value of the static model significantly. For example, when the investor's  expected terminal wealth is $12$ (or equivalently, the expected target return is $20\%$) and the confidence level is $95\%$, the correspondent CVaR is $2.849$ if he implements the buy and hold static portfolio policy and the CVaR is only $0.208$ if he implements the dynamic mean-CVaR portfolio policy. Figure \ref{fig_efficient} plots the mean-CVaR efficient frontiers of the buy-and-hold (BnH) policies and our dynamic portfolio policy (Dyn). We can see that the efficient frontiers of the buy-and-hold policy are more sensitive than the ones generated by the dynamic mean-CVaR policy when the confidence level $\beta$ increases.

\begin{figure}
  \centering
  \includegraphics[width=300pt]{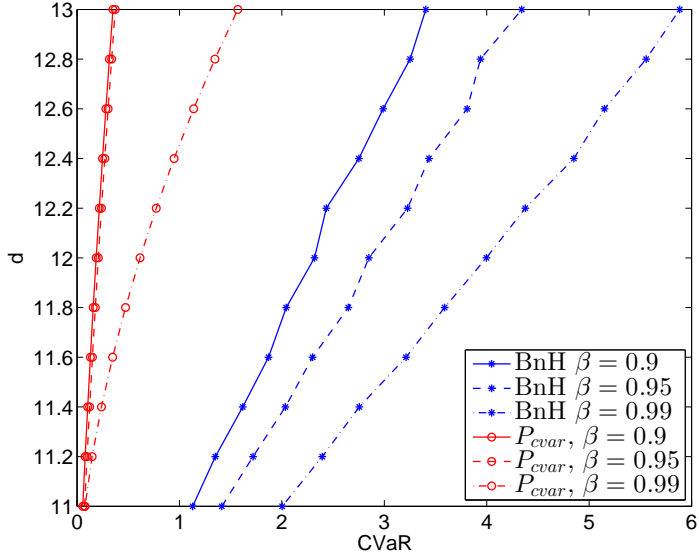}\\
  \caption{The mean-CVaR efficient frontiers of Example \ref{exam_cvar} generated by the buy-and-hold policy and the dynamic policy}\label{fig_efficient}
\end{figure}

\end{exam}

%*******************************************************************************

\section{Conclusion}\label{Conclusion}

We have investigated in this paper two long-standing challenges in dynamic portfolio selection, the dynamic mean-LPM and dynamic mean-CVaR portfolio optimization problems, and have solved both completely. By adding a limited funding level on the terminal wealth, we ensure the well-posedness of the two problems, which further enables us to adopt the martingale approach in characterizing the solution. We have proved that, under some mild conditions,  Lagrange multipliers always exist for the static hedging equations, which is the key in adopting such a martingale approach. When the market opportunity set is deterministic, we can achieve analytical portfolio policies of these problems. Our examples show that the dynamic mean-LPM portfolio policy performs better than the well known mean-variance portfolio policy with respect to management of the downside risk. Compared with the static buy-and-hold mean-CVaR portfolio policy, the dynamic portfolio policy can reduce the CVaR level significantly. Our dynamic mean-downside-risk portfolio shows some prominent features, e.g., implementing such a policy can control the CVaR value at a very low level even when the expected return is set at a high level. However, the price of using these portfolio policies could be also quite high. Usually, a dynamic mean-LPM or mean-CVaR policy requires to short a large amount of some assets in the portfolio. Thus, it would be more realistic to impose a no-shorting constraint in the dynamic mean-LPM and dynamic mean-CVaR models, which deserves our future endeavors.

\section*{Appendix: Proof of Lemma \ref{lem_lag}}

\proof In the following proof, we use $v(\cdot)$ to denote the optimal value of problem $(\cdot)$. Since the optimal solution of problem ($\mathcal{B}$) is a feasible solution of problem $(L(\lambda_1,\lambda_2))$, we have a weak duality relationship, $v(L(\lambda_1,\lambda_2))\leq v(\mathcal{B})$, for any $\lambda_1 \in \R_+$ and $\lambda_2 \in \R$. On the other hand, if $Y^*\in C$ solves problem ($\L(\lambda_1^*,\lambda_2^*)$) and $Y^*$ satisfies $\E[Y^*]\geq b$ and $\E[ZY^*]=a$, then $Y^*$ is a feasible solution of $(\B)$, which implies $v(\B)\leq v(\L(\lambda_1^*,\lambda_2^*))$. Together with the weak duality relationship, we have $v(\L(\lambda_1^*,\lambda_2^*))=v(\B)$, which further implies $\lambda_1^*(\E[Y]-b)=0$ and $\lambda^*_2(\E[ZY]-a)=0$. That is to say, $Y^*$ solves the problem $(\B)$.

Now, we prove the other direction. Let $Y^*$ be the solution of problem ($\mathcal{B}$) and $J^*=v(\mathcal{B})$. We construct the epigraph set of problem $(\B)$ as
\begin{align}
\mathcal{O}&:=\big\{(\kappa_1,\kappa_2,\kappa_3)^{\prime}\in \R^3~|~\exists~Y\in C,~ b-\E[Y]\leq \kappa_1,\notag\\
&~~~~~~~~~~~~~\E[ZY]-a=\kappa_2, ~\kappa_3\geq \E[f(Y)]~\big\}.\label{def_O}
\end{align}
Obviously, due to the convexity of $f(\cdot)$, set $\O$ is a convex set in $\R^3$. We construct another set $\mathcal{M}:=\big\{(0,0,J)\in \R^3~|~J<J^*\}$, which is also convex. We have $\O \bigcap \M=\emptyset$. If $\O \bigcap \M \not=\emptyset$, there exists $(0,0, \hat{J}) \in \O \bigcap \M$. Since $(0,0,\hat{J})\in \M$, we have $\hat{J}<J^*$. Similarly, $(0,0, \hat{J})\in \O$ implies that there exists $\hat{Y}\in C$ such that $\E[\hat{Y}]\geq b$, $\E[Z\hat{Y}]=a$ and $\E[f(\hat{Y})]\leq \hat{J}<J^*$. That is to say, $\hat{Y}$ is the solution of problem $(\B)$ with a smaller objective value, which contradicts the optimality of $Y^*$. Since both $\O$ and $\M$ are convex sets and do not intersect each other, by using the Separating Hyperplane Theorem \cite{Rockafellar:1996}, there exist $\epsilon$ and $({\phi}_1, {\phi}_2, {\phi}_3)\not =(0,0,0)$ such that, for any $(\kappa_1,\kappa_2,\kappa_3)\in \O$ and $(0,0,J)\in \M$,
\begin{align}
&\phi_1\kappa_1+\phi_2\kappa_2+\phi_3\kappa_3\geq \epsilon,\label{lem_lag_ineq1}\\
&\phi_3 J\leq \epsilon.\label{lem_lag_ineq2}
\end{align}
From the definition in (\ref{def_O}), we must have $\phi_1\geq 0$ and $\phi_3\geq0$. Otherwise, $\phi_1\kappa_1+\phi_3\kappa_3$ is unbounded from below over $\O$ ($\kappa_1$ and $\kappa_3$ can go to infinity in set $\O$), which contradicts (\ref{lem_lag_ineq1}). Condition (\ref{lem_lag_ineq2}) implies that $\phi_3 J \leq \epsilon$ for all $J<J^*$, and we thus have $\phi_3J^*\leq \epsilon$. Together with (\ref{lem_lag_ineq1}), we have
\begin{align}
&\phi_3 f(Y)+\phi_1( b-\E[Y])+\phi_2(\E[ZY]-a)\geq \epsilon \geq \phi_3 J^*,\label{lem_lag_inequ3}
\end{align}
for any $Y\in C$. Now we first assume $\phi_3>0$. Dividing both sides of (\ref{lem_lag_inequ3}) by $\phi_3$ gives rise to
\begin{align}
f(Y)+\bar{\lambda}_1( b-\E[Y])+\bar{\lambda}_2(\E[ZY]-a)\geq J^*,\label{lem_lag_inequ4}
\end{align}
for any $Y\in C$, where $\bar{\lambda}_1=\phi_1/\phi_3$ and $\bar{\lambda}_2=\phi_2/\phi_3$. Together with the weak duality relationship, the inequality in (\ref{lem_lag_inequ4}) implies that $v(\B)=v(\L(\bar{\lambda}_1,\bar{\lambda}_2))$ and $\bar{\lambda}_1(\E[Y]-b)=0$. Thus, $Y^*$ solves problem ($\L(\bar{\lambda}_1,\bar{\lambda}_2)$), which completes the proof for the case when $\phi_3>0$.

Now, we show that $\phi_3\not=0$. If $\phi_3=0$, the inequality in (\ref{lem_lag_inequ3}) becomes,
\begin{align}
\phi_1( b-\E[Y])+\phi_2(\E[ZY]-a)\geq 0, \label{lem_lag_inequ5}
\end{align}
for any $Y\in C$. Let $\bar{Y}$ be some interior feasible solution of $(\B)$ and condition (\ref{lem_lag_inequ5}) becomes $\phi_1( b-\E[\bar{Y}])\geq 0$. Due to the strict feasibility, we have $b-\E[\bar{Y}]<0$, which implies $\phi_1=0$. Note that $(\phi_1,\phi_2,\phi_3)\not=(0,0,0)$, thus, $\phi_2\not=0$. Thus, the condition in (\ref{lem_lag_inequ5}) becomes $\phi_2(\E[ZY]-a)\geq 0$ for all $Y\in C$, which is impossible. Note that there is a strictly interior feasible solution $\bar{Y}\in C$ such that $\phi_2(\E[Z\bar{Y}]-a)=0$. That is to say, in the neighborhood of $\bar{Y}$, we can always find $\tilde{Y}\in C$ such that $\phi_2(\E[Z\tilde{Y}]-a)<0$.\footnote{A strictly interior feasible point is also called a relative interior point of the feasible set. Let $F_B$ be the feasible set of problem $(\B)$. For a given point $\bar{Y}$, if we can find an open ball $O_B$ centered at $\bar{Y}$ such that $F_B \bigcap O_B \subset F_B$, then $\bar{Y}$ is a strictly feasible interior point.} Thus, we can conclude that $\phi_3\not=0$ and thus complete our proof.
\endproof

\section*{Appendix: Lemma \ref{lem_expectation}}
\begin{lem}\label{lem_expectation}
Let $Y$ be a random variable that follows the normal distribution with mean $\mu$ and variance $v^2$, respectively. Then, we have
\begin{align}
\E[e^{aY}\cdot \1_{Y \leq d}]&=\exp\left(a\mu+\frac{a^2v^2}{2}\right) \Phi\left(\frac{d-\mu}{v}-av\right), \label{lem_expectation1}
%\E[e^{aY}\cdot \1_{Y \geq d}]&=\exp\left(a\mu+\frac{a^2v^2}{2}\right) \left(1-\Phi\left(\frac{d-\mu}{v}-av\right)\right)\label{lem_expectation2}
\end{align}
where $\Phi(\cdot)$ is the cumulative distribution function of standard normal random variable.
\end{lem}
\proof
Let $Z=(Y-\mu)/v$. Then $Z$ follows the standard normal distribution and
\begin{align*}
\E[e^{aY}\1_{Y\leq d}]&=\E[e^{a(zv+\mu)}\1_{zv+\mu\leq d} ]\\
&=\frac{1}{\sqrt{2\pi}}\int_{-\infty}^{\frac{d-\mu}{v}} \exp\left(-\frac{(z^2-2azv-2a\mu}{2}\right) dz\\
&=\frac{1}{\sqrt{2 \pi}} \exp(\frac{2a\mu+a^2v^2}{2})\int_{-\infty}^{\frac{d-\mu}{v}} \exp \left(-\frac{(z-av)^2}{2}  \right)dz\\
&=\exp\left(a\mu+\frac{a^2v^2}{2}\right) \Phi\left( \frac{d-\mu}{v}-av\right),
\end{align*}
which is exactly (\ref{lem_expectation1}).
\endproof

%******************************************************************************
%****************************  bib  *******************************************


\begin{thebibliography}{10}

\bibitem{Alexander:2006}
{\sc S. Alexander, T. F. Coleman and Y. Li}, {\em Minimizing CVaR and VaR for portfolio of derivatives}, J. Banking Finance, 30(2006), pp. 583-605.

\bibitem{Andersson:2001}
{\sc F. Andersson, H. Mausser, D. Rosen and S. Uryasev}, {\em Credit risk optimization with Conditional Value-at-Risk criterion}, Math. Program., Series B, 89(2001), pp. 273-291.

\bibitem{Artzner:1999}
{\sc P. Artzner, F. Delbaen, J. M. Eber and D. Heath}. {\em Coherent measure of risk},  Math. Finance, 9(1999), pp. 203-228.

\bibitem{Basak:2010}
{\sc S. Basak and G. Chabakauri}, {\em  Dynamic mean-variance asset allocation},
Rev. Financ. Studies, 23(2010), pp. 2970-3016.


\bibitem{Basak:2001}
{\sc S. Basak and A. Shapiro}, {\em Value-at-Risk-based risk management: Optimal policies and asset prices}, Rev. Financ. Studies, 14(2001), pp. 371-405.

%\bibitem{Bawa:1975}
%Bawa, V.S. 1975. Optimal rules for ordering uncertian prospects, {\it Journal of Financial Economics}, {\bf 2}, 95-121.

\bibitem{Bawa:1977}
{\sc V. S. Bawa and E. B. Lindenberg}, {\em Capital market equilibrium in a mean-Lower Partial Moment framework}, J. Finacial Economics, 5(1977), pp. 189-200.

\bibitem{Bielecki:2005}
{\sc T. Bielecki, H. Q. Jin, S. R. Pliska and X. Y. Zhou}, {\em Continuous-time mean-variance portolio selection with bankrupcy prohibition},  Math. Finance, 15(2005), pp. 213-244.

\bibitem{Bjork:2012}
{\sc T. Bj\"{o}rk, A. Murgoci and X. Y. Zhou}, {\em Mean-variance portfolio optimization with state-dependent risk aversion}, Math. Finance,  DIO: 10.1111/j.1467-9965.2011.00515.x.

\bibitem{CLWZ:2011}
{\sc X. Y. Cui, D. Li, S. Y. Wang and S. S. Zhu}, {\em Better than dynamic mean-variance: Time inconsistency and free cash flow
stream}, Math. Finance, 22(2012), pp. 346-378.

\bibitem{Chiu:2012}
{\sc M. C. Chiu, H. Y. Wong and D. Li}, {\em Roy's safety-first portfolio principle in financial risk management of disastrous events}, Risk Analysis, 32(2012), pp. 1856-1872.

\bibitem{Dembo:1999}
{\sc R. Dembo and D. Rosen}, {\em The practice of portfolio replication: a practical overview of forward and inverse problems},  Ann. Oper. Res., 85(1999), pp. 267-284.


\bibitem{Fabian:2006}
{\sc C. I. F\'{a}bi\'{a}n}, {\em Handling CVaR objectives and constraints in two-stage stochastic models}, European J. Oper. Res., 191(2006), pp. 888-911.

\bibitem{Fabian:2007}
{\sc C. I. F\'{a}bi\'{a}n and Z. Szoke}, {\em Solving two-stage stochastic programming problems with level decomposition}, Comput. Manag. Sci., 4(2007), pp. 313-353.


\bibitem{Fishburn:1977}
{\sc F. C. Fishburn}, {\em Mean-risk analysis with risk associated with below-target returns}, Amer. Econoc. Rev., 67(1977), pp. 116-126.


\bibitem{Frittelli:2000}
{\sc M. Frittelli}, {\em The minimal entropy martingale measure and the valution problem in incomplete markets}, Math. Finance, 10(2000), pp. 39-52.



\bibitem{Gundel:2008}
{\sc A. Gundel and S. Weber}, {\em Utility maximization under a shortfall risk constraint},  J. Math. Econom., 44(2008), pp. 1126-1151.

\bibitem{Hibibi:2006}
{\sc N. Hibiki}, {\em Multi-period stochastic optimization models for dynamic asset allocation}, J. Banking \& Finance, 30(2006), pp. 365-390.


\bibitem{Jin:2005}
{\sc H. Q. Jin, J. A. Yan and X. Y. Zhou}, {\em Continuous-time mean-risk portfolio selection}, Ann. Henri Poincar\'{e}, 41(2005), pp. 559-580.

\bibitem{Jin:2008}
{\sc H. Q. Jin and X. Y. Zhou}, {\em Convex stochastic optimization problem arising from portfolio selection}, Math. Finance, 18(2008), pp. 171-183.



\bibitem{Karatzas:1998}
{\sc I. Karatzas and S. E. Shreve}, {\em Methods of Mathematical Finance}, New York: Springer-Verlag, 1998.

\bibitem{Karoui:1997}
{\sc N. EL Karoui, S. Peng and M. C. Quenez}, {\em Backward stochastic differential equations in finance}, Math. Finance, 7(1997), pp. 1-71.

\bibitem{Konno:2002}
{\sc H. Konno, H. Waki and A. Yuuki}, {\em Portfolio optimization under lower partial risk measures}, Asia Pac. Financ. Market, 9(2002), pp. 127-140.


\bibitem{LiNg:2000}
{\sc D. Li and W. L. Ng}, {\em Optimal dynamic portfolio selection:
Multiperiod mean-variance formulation},  Math. Finance, 10(2000), pp. 387-406.

\bibitem{LiZhou:2001}
{\sc X. Li, X. Y. Zhou and A. E. B. Lim}, {\em Dynamic mean-variance
portfolio selection with no-shorting constraints}, SIAM J. Control Optim, 40(2001), pp. 1540-1555.


\bibitem{Lim:2002}
{\sc A. E. B. Lim and X. Y. Zhou}, {\em Mean-variance portfolio
selection with random parameters in a complete market}, Math. Oper. Res., 27(2002), pp. 101-120.

\bibitem{Markowitz:1952}
{\sc H. M. Markowitz}, {\em Portfolio Selection}, J. Finance,  7(1952), pp.77-91.


\bibitem{Ogryczak:2002}
{\sc W. Ogryczak and A. Ruszczy\'{n}ski}, {\em Dual stochastic dominance and related mean-risk models}, SIAM J. Optim., 13(2002), pp. 60-78.


\bibitem{Pliska:1986}
{\sc S. R. Pliska}, {\em A stochastic calculus model of continuous trading: Optimal portfolios optimization}, Math. Oper. Res., 11(1986), pp. 371-384.


\bibitem{Rockafellar:1996}
{\sc R. T. Rockafellar}, {\em Convex Analysis}, Princeton University Press, 1996.

\bibitem{Rockafellar:2000}
{\sc R. T. Rockafellar and S. Uryasev}, {\em Optimization of conditional value-at-risk}, J. Risk, 2(2000), pp. 21-41.


\bibitem{Rockafellar:2002}
{\sc R. T. Rockafellar and  S. Uryasey},  Conditional Value-at-Risk for general loss distributions, {\it Journal of Banking and Finance}, {\bf 26} 1443-1471.

\bibitem{Roy:1952}
{\sc A. D. Roy}, {\em Safety first and the holding of assets}, Econometrica, 20(1952), pp. 431-449.


\bibitem{Schweizer:1996}
{\sc M. Schweizer}, {\em Aproximation pricing and the variance-optimal martingale measure}, Ann. Probab., 24(1996), pp. 206-236.

\bibitem{Yiu:2004}
{\sc K. F. C. Yiu}, {\em Optimal portfolios under a value-at-risk constraint},  J. Econom. Dynam. Control, 28(2004), pp. 1317-1334.


\bibitem{ZhouLi:2000}
{\sc X. Y. Zhou and D. Li}, {\em  Continuous time mean-variance portfolio
selection: A stochastic LQ framework}, Appl. Math. Optim., 42(2000), pp. 19-33.

\bibitem{Zhu:2009}
{\sc S. S. Zhu and M. Fukushima}, {\em  Worst-case conditional Value-at-Risk with application to robust portolio management},  Oper. Res., 57(2009), pp. 1155-1168.


\bibitem{Zhu:2004}
{\sc S. S. Zhu, D. Li and S. Y. Wang}, {\em Risk control over
bankruptcy in dynamic portfolio selection: A generalized mean-variance formulation}, IEEE Trans. Automat. Control, 49(2004), pp. 447-457.

\bibitem{ZhuLiWang:2009}
{\sc S. S. Zhu, D. Li and S. Y. Wang}, {\em Robust portfolio selection under downside risk measures}, Quant. Finance, 9(2009), pp. 869-885.


\end{thebibliography}
\end{document}